\documentclass[a4paper]{article}

\usepackage[utf8]{inputenc}
\usepackage[OT4]{fontenc}

\usepackage{amsmath, amssymb, amsthm}
\usepackage{thmtools}
\usepackage{thm-restate}

\usepackage{geometry}
\usepackage{microtype}
\usepackage{xspace}
\usepackage{enumerate}
\usepackage{enumitem}
\usepackage{array, booktabs, makecell, multirow, tabularx}

\usepackage[ruled]{algorithm}
\usepackage[noend]{algpseudocode}
\algnewcommand{\LineComment}[1]{\State \(\triangleright\) #1}

\usepackage{graphicx} \DeclareRobustCommand{\COMPASSicon}{\raisebox{-0.15em}[0pt][0pt]{\includegraphics[height=1em]{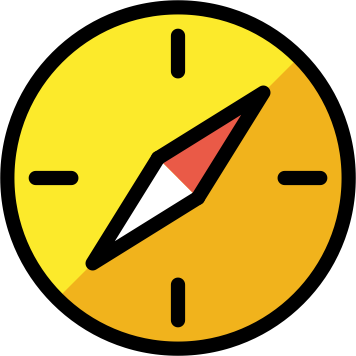}}\hspace{0.15em}}
\usepackage[dvipsnames]{xcolor}
\definecolor{darkblue}{rgb}{0,0.08,0.45}
\usepackage{pgf, tikz}
\usepackage{subcaption}

\usepackage[noadjust]{cite}
\usepackage[hypertexnames=false,colorlinks=true,
            urlcolor=Blue,citecolor=Green,linkcolor=BrickRed]{hyperref}
\usepackage[capitalise]{cleveref}

\usepackage{verbatim}
\usepackage{comment}
\usepackage{units}
\usepackage{todonotes}
\usepackage[title]{appendix}

\newcolumntype{P}[1]{p{#1}<{\raggedright}}

\makeatletter
\def\@seccntformat#1{\@ifundefined{#1@cntformat}%
   {\csname the#1\endcsname\quad}%
   {\csname #1@cntformat\endcsname}}
\let\oldappendix\appendix
\renewcommand\appendix{%
    \oldappendix
    \newcommand{\section@cntformat}{\appendixname~\thesection\quad}
}
\makeatother

\begin{document}

\begin{center}
    {\Large\bfseries Learning to Transfer School Attendance Zoning across Cities\par}
    \vspace{0.75em}
    {\large Ruiya Zhou\textsuperscript{1} \quad Zhanzhan Zhao\textsuperscript{1,2,*}\par}
    \vspace{0.25em}
    {\small \textsuperscript{1}School of Humanities and Social Science, The Chinese University of Hong Kong, Shenzhen, Shenzhen, China\par}
    {\small \textsuperscript{2}School of Artificial Intelligence, The Chinese University of Hong Kong, Shenzhen, Shenzhen, China\par}
    {\small \textsuperscript{*}Corresponding author: Zhanzhan Zhao (zhanzhanzhao@cuhk.edu.cn)\par}
\end{center}

\begin{abstract}
Designing multi-school attendance zones requires selecting constraint settings that balance educational equity, accessibility, and school capacity.
Repeating this process for each new city can require substantial local calibration and computational search.
We introduce COMPASS, a parameter-to-policy framework that transfers knowledge from previous zoning searches to unseen cities through supervised prediction.
Using search-derived parameter configurations as offline supervision, COMPASS learns to predict interpretable zoning-control parameters from urban representations and zoning-difficulty signals.
A downstream mixed-integer linear program converts these parameters into explicit zoning plans, with conditional parameter repair when the initial solve is unsuccessful.
We evaluate parameter prediction, zoning-outcome recovery, and computational efficiency through five-fold cross-validation across 3,000 US cities, complemented by a case study of Atlanta.
COMPASS achieves a 99.80\% solver-feasible success rate and a \(49.14\times\) median deployment-time speedup relative to repeated city-specific parameter search.
Further analysis shows that transfer performance varies across urban contexts and policy dimensions: structurally difficult cities receive the largest feasibility gains, while enrollment balance remains the hardest outcome to recover.
These findings demonstrate how cross-city learning can reduce repeated search while retaining explicit zoning controls and spatial plans for local inspection and adaptation.
\end{abstract}


\section{Introduction}
\label{sec:introduction}

Geographic boundaries help determine which public schools families can access and how educational opportunities are distributed across urban space.
Because neighborhood-based assignment ties school access to residential location, school composition often reflects patterns of residential segregation.
Attendance boundaries can reinforce these patterns or create opportunities for greater integration \cite{monarrez2023}.
Improving demographic balance, however, may lengthen students' journeys or shift enrollment toward schools with limited capacity, whereas prioritizing proximity and balanced enrollment may restrict opportunities for integration.
These trade-offs shape how benefits and burdens are distributed across families.
Deciding which compromises are acceptable requires reconciling different stakeholder priorities, making school zoning both a computational problem and a consequential policy decision \cite{castro2024}.

Computational optimization provides a systematic way to examine these trade-offs.
Objectives and constraints can represent requirements concerning accessibility, enrollment, capacity, demographic composition, school quality, and geographic contiguity \cite{ferland1990,chen2023tradeoffs}.
Applying such a formulation requires concrete control parameters, including the number of zones, permitted enrollment deviations, allowable coverage distances, and demographic bounds.
These parameters translate policy requirements into operational settings, defining which zoning plans are admissible and which compromises the optimizer can consider.
Obtaining a suitable plan therefore requires both solving the optimization problem and configuring its controls appropriately.

Selecting these controls is difficult because their effects interact.
Tight demographic bounds may conflict with the spatial distribution of students, while capacity and contiguity requirements further restrict the feasible partitions.
A configuration whose individual settings appear reasonable may therefore admit no feasible plan when the constraints are imposed jointly.
Relaxing these settings can restore feasibility but may compromise the intended zoning outcomes.
Finding a suitable configuration can consequently require repeated cycles of parameter selection, optimization, evaluation, and adjustment, supported by interactive decision tools or computational search \cite{holloway1975,ferland1990,desjardins2007,biswas2020}.

Applying the same framework to a new city creates a further configuration challenge.
Differences in spatial structure, school supply, student demand, and demographic composition can make previously selected parameter values unsuitable.
Repeating the full calibration or search process independently for every city leaves potentially useful experience from earlier applications underused.
This motivates transferring knowledge about parameter configuration across cities.
The question is whether the relationship between urban conditions and suitable control settings can be learned from previous searches and used to propose parameters for a target city.
Such transfer could reduce repeated configuration work and make computational zoning methods easier to apply across cities, while retaining local data, policy goals, and explicit controls over zoning requirements.

\begin{figure}[t]
\centering
\IfFileExists{figures/ZoneOptimization_Figure_6.pdf}{\includegraphics[width=\textwidth,keepaspectratio]{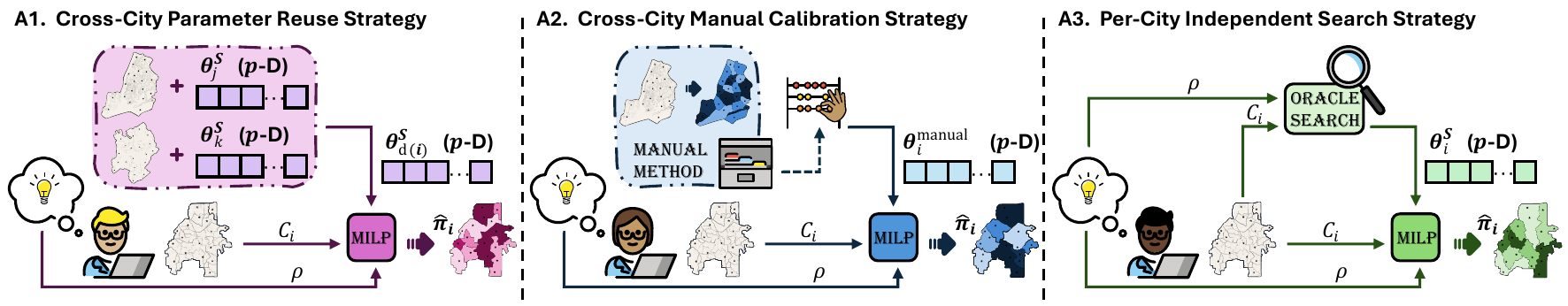}}{\fbox{\parbox[c][3.25cm][c]{0.88\textwidth}{figures/ZoneOptimization_Figure_6.pdf}}}
\caption{(\textit{A}) Three conventional strategies are used to generate school-zoning control parameters: (\textit{A1}) directly reusing parameters calibrated for another similar city, (\textit{A2}) manually calibrating parameters for the target city, and (\textit{A3}) performing an independent parameter search for the target city.}
\label{fig:ZoneOptimization_Figure_6}
\end{figure}

Existing approaches provide several ways to obtain these target-city controls.
One strategy directly reuses parameters from similar cities \textbf{(Figure~\hyperref[fig:ZoneOptimization_Figure_6]{\ref*{fig:ZoneOptimization_Figure_6}.A1})}, requiring source-city matching and evaluation of whether the borrowed settings remain suitable.
A second calibrates parameters manually according to local conditions \textbf{(Figure~\hyperref[fig:ZoneOptimization_Figure_6]{\ref*{fig:ZoneOptimization_Figure_6}.A2})} \cite{holloway1975interactive,ferland1990decision}, potentially involving repeated adjustments and optimization runs.
A third conducts an independent parameter search for each new city \textbf{(Figure~\hyperref[fig:ZoneOptimization_Figure_6]{\ref*{fig:ZoneOptimization_Figure_6}.A3})} \cite{desjardins2007heuristic,biswas2020incorporating}, automating configuration exploration but repeating its computational cost.
These strategies differ in their reliance on human input and computation, but none explicitly learns a reusable mapping from urban conditions to zoning-control parameters.

Learning-based approaches offer a route to exploiting shared structure across instances, including reinforcement learning for school districting \cite{zhang2024deep} and structured learning coupled with optimization for geographical districting \cite{ahmed2024districtnet}.
For school attendance zoning, an important design requirement is to retain explicit controls through which decision-makers can inspect and adjust how policy requirements enter the optimization \cite{papadakis2024explainable}.
This motivates combining cross-city learning at the parameter level with constrained plan construction.
Its effectiveness must also be assessed beyond prediction accuracy and runtime: suitable predictions must enable feasible plans and recover the intended zoning outcomes.
Understanding how this performance varies across urban contexts and policy dimensions is therefore essential to establishing the benefits and limits of transfer.

These practical difficulties motivate the following two research questions in our paper:
\begin{center}
    \phantomsection\label{rq:q1}\textbf{Q1 (Methodological).} \textit{How to enable a resource-efficient policy transfer via an interpretable framework?}\\
    \phantomsection\label{rq:q2}\textbf{Q2 (Empirical).} \textit{How transferability varies across urban contexts and policy dimensions?}
\end{center}

The answer to question \hyperref[rq:q1]{\textbf{Q1}} \textbf{(Methodological)} is our COMPASS, a parameter-to-policy framework compressing parameter search into supervised prediction.
Taking the optimization-ready urban representations and the school-zoning difficulty of cities as inputs, COMPASS uses existing parameter configurations as offline supervision, so that it learns to generate interpretable zoning-control parameters adapted to target cities.
When deployed to an unseen city, COMPASS generates the required control parameters within one forward path, and a downstream mixed-integer linear programming (MILP) model converts these parameters into a constraint-satisfying school-zoning plan.
This parameter-interpretable pipeline reduces both human effort and computational cost, achieving resource-efficient policy transfer.

We answer question \hyperref[rq:q2]{\textbf{Q2}} \textbf{(Empirical)} with two key observations by analyzing our COMPASS method over 3,000 US cities.
According to our analysis, school-zoning transferability is shaped by two factors: the urban contexts of target cities and the policy preferences.
In terms of urban contexts, we find that cities with higher structural difficulties exhibit the worst transferability.
This type of cities receive the most feasibility improvements from COMPASS.
By contrast, large-scale cities only have the heaviest computational burdens, which bring no increase in zoning difficulties or the infeasibility rate.
Across policy dimensions, student balance remains the hardest to recover, while transferability of other dimensions varies with overall policy preferences.

Our contributions are summarized as follows:
\begin{itemize}
    \item We propose COMPASS, an interpretable parameter-to-policy framework \textbf{(Section \ref{subsec:framework_overview})}, and further construct an optimization-ready urban representation \textbf{(Section \ref{subsec:optimization_ready_urban_representation})} and a difficulty-aware parameter generation \textbf{(Section \ref{subsec:difficulty_aware_parameter_generation})}. These designs together form a complete cross-city chain from city conditions to interpretable control vectors and then to school-zoning plans.
    \item Through 5-fold cross-city experiments on our dataset of 3,000 US cities, we validate COMPASS in terms of its parameter-prediction accuracy, policy-recovery effectiveness, and computational efficiency \textbf{(Section \ref{subsec:experimental_evaluation})}, when facing unseen target cities. We further demonstrate the end-to-end pipeline of COMPASS in practice through a specific case study \textbf{(Section \ref{subsec:case_study_atlanta})}.
    \item We identify systematic patterns in how the school-zoning transferability varies with urban contexts and policy dimensions \textbf{(Section \ref{subsec:transferability_patterns})}. Based on these differences, we further clarify responsibility boundary of COMPASS in policy implementation \textbf{(Section \ref{subsec:policy_implementation})}.
\end{itemize}

\section{Related Work}
\label{sec:related_work}

\paragraph{School zoning.} 
School attendance zoning has been a longstanding challenge for cities \cite{castro2024drawn}.
To achieve educational equity for student populations, an attendance-zone partition must take multiple factors into consideration including boundary shape, residential distribution \cite{saporito2017irregularly, monarrez2023school}, travel burdens, and school quality \cite{asson2024rezoning}.
These constraints make school zoning a multi-objective problem rather than a purely geometric partition \cite{chen2023exploring}.
Facing this problem, desJardins et al. \cite{desjardins2007heuristic} and Biswas et al. \cite{biswas2020incorporating} generated alternative zoning plans by performing independent heuristic or domain-aware spatial optimization for each district.
More broadly, other researchers employed learning-based \cite{bengio2021machine, kadioglu2010isac} and amortized optimization \cite{amos2023tutorial,  sambharya2024learning, bertsimas2022online} to accelerate future problem solving by reusing computation from previous problem instances and producing reusable learned outputs.
However, these general-purpose methods lack school-zoning-specific structure, where we tackle this limitation and provide interpretable parameterization to support local policy design.

\paragraph{Policy transfer.}
Public policies are often transferred across different cities due to their societal value or administrative requirements \cite{mccann2013multidisciplinary}.
However, a direct transfer without target city knowledge can be arbitrary and ineffective \cite{blanc2023spatial}.
To bridge this gap, the literature introduces various domain adaptation strategies, which specifically focus on the heterogeneous parts between cities \cite{canitez2020transferring, noori2023transplanting}.
In detail, these studies quantify urban heterogeneity through either quantitative urban morphology \cite{fleischmann2021measuring, fleischmann2022methodological}, or urban-region embeddings \cite{li2023urban, zhang2020multiview}.
Despite these advances, current solutions exhibit limitations in application domains.
In this work, we design a systematic representation specifically for school zoning to support policy transfer in this context.

\section{Problem Formulation}
\label{sec:problem_formulation}

We formulate school attendance zoning as a parameterized graph-partitioning problem.
To formalize this problem mathematically, we define \(C_i=(G_i,\mathbf a_i)\) as the instance of city \(i\), where \(G_i=(\mathcal V_i,\mathcal E_i)\) is the adjacency graph of geographic assignment units and \(\mathbf a_i\) contains the policy-relevant attributes. 
The zoning task then divides these \(n_i=|\mathcal V_i|\) units into \(K_i\) zones, where \(1\leq K_i\leq n_i\) and \(\mathcal V_{ik}\subseteq\mathcal V_i\) denote the units assigned to zone \(k\), yielding the zoning plan \(\boldsymbol{\pi}_i=\left(\mathcal V_{i1},\ldots,\mathcal V_{iK_i}\right)\).
At the structural level, a valid division must cover every geographic unit exactly once, contain no empty zones, and preserve geographic contiguity within every zone.
The set of zone partitions satisfying the above requirements is formally defined as
\begin{equation}
\label{eq:set_of_all_zoning_plans}
\mathcal{Z}_{K_i}(G_i) = \left\{(\mathcal{V}_{i1}, \dots, \mathcal{V}_{iK_i})\;\middle|\;\biguplus_{k=1}^{K_i} \mathcal{V}_{ik} = \mathcal{V}_i,~\mathcal V_{ik}\neq\varnothing,~\operatorname{Connect}(
G_i[\mathcal V_{ik}])=1~\forall k\right\}.
\end{equation}
Beyond structural validity, however, school zoning, ultimately aiming at a real policy decision, must account for the city attributes in \(\mathbf a_i\), including student demand, demographic composition, school capacity, and school quality.

To translate these policy considerations into quantitative controls, we introduce a \(p\)-dimensional control vector for the zoning optimization model:
\begin{equation}
\label{eq:zoning_parameters}
\boldsymbol{\theta}_i=\Big(K_i,\tau_i,\kappa_i,\eta_i,s_i^{-},s_i^{+},\left\{\ell_{ig},u_{ig}\right\}_{g\in\mathcal G}\Big)\in\mathbb R^p,
\end{equation}
where \(\mathcal G=\{\mathrm{FRL},\mathrm{FG},\mathrm{DI},\mathrm{SQ}\}\), and \(p=14\) in this study.
Table \ref{tab:control_vector} summarizes the fields, symbols, counts, and definitions of each control.
These controls must lie in the following common admissible domain:
\begin{equation}
\label{eq:admissible_domain}
\Theta:=\left\{\boldsymbol{\theta}\in\mathbb{R}^p : K\in\mathbb Z_{\geq1},~s^{-},s^{+}\in\mathbb Z_{\geq0},~ \tau,\eta,\kappa,\ell_{g},u_{g}\geq0,~\ell_{g}\leq u_{g}, \forall g\in\mathcal G\right\}.
\end{equation}

\begin{table}[t]
\centering
\caption{Components of the school-zoning control vector.}
\label{tab:control_vector}
\setlength{\tabcolsep}{5pt}
\begin{tabular*}{\textwidth}{
@{\extracolsep{\fill}}lccl}
\toprule
Field & Symbol & Count & Definition \\
\midrule
Zone structure & \(K_i\) & 1 & Number of school zones in the configuration\\
Enrollment balance & \(\tau_i\) & 1 & Permitted enrollment deviation across zones\\
Spatial accessibility & \(\kappa_i\) & 1 & Multiplier determining the allowable coverage distance\\
Capacity shortage & \(\eta_i\) & 1 & Maximum capacity shortage permitted within a zone\\
School distribution & \(s_i^-,s_i^+\) & 2 & Deviations from the average number of schools per zone\\
Equity bounds & \((\ell_{ig},u_{ig})_\mathrm{FRL,FG,DI}\) & 6 & Bounds on socioeconomic and demographic composition\\
School quality bounds & \((\ell_i,u_i)_{\mathrm{SQ}}\) & 2 & Lower and upper bounds on school quality\\
\bottomrule
\end{tabular*}
\vspace{2pt}
\raggedright
\footnotesize
FRL: free or reduced-price lunch;
FG: focal group;
DI: diversity index;
SQ: school quality.
\end{table}

However, \(\boldsymbol{\theta}_i\in \Theta\) only guarantees parameter-level validity; the corresponding city-specific zoning problem may still have no feasible solution.
The feasibility of a parameter-valid control vector for a particular city is determined by solving a MILP for city instance \(C_i\) with the candidate \(\boldsymbol{\theta}_i\in\Theta\) and a fixed policy specification \(\boldsymbol{\rho}\).
Under this configuration, the MILP search for a structurally valid zoning plan \(\boldsymbol{\pi} \in\mathcal{Z}_{K_i}(G_i)\), which also satisfies enrollment balance, capacity shortage, equity, school distribution, school quality, spatial accessibility, and contiguity requirements induced by  \(\boldsymbol{\theta}_i\) and \(C_i\).
If no such plan exists, this candidate vector is regarded as infeasible.
Otherwise, the MILP selects a feasible zoning plan as the current optimal solution:
\begin{equation}
\label{eq:zoning_milp_return}
\boldsymbol{\pi}_i^{\mathrm{MILP}}(\boldsymbol{\theta}_i,\boldsymbol{\rho})\in\arg\min_{\boldsymbol{\pi}\in\mathcal{Z}_{K_i}(G_i)}J_i(\boldsymbol{\pi};\boldsymbol{\rho}),
\end{equation}
where \(J_i\) evaluates each feasible plan according to the fixed policy specification, and \(\boldsymbol{\pi}_i^{\mathrm{MILP}}\) denotes the selected one.
In Section \ref{sec:compressing_parameter_search_into_supervised}, we introduce COMPASS to address the remaining problem: generating a city-specific, policy-interpretable and MILP-feasible control vector for an unseen city in a single step.

\section{Compressing Parameter Search into Supervised Prediction}
\label{sec:compressing_parameter_search_into_supervised}

In this section, we answer the methodological question \hyperref[rq:q1]{\textbf{Q1}} by introducing our COMPASS as follows. 
First, we construct the overall COMPASS framework \textbf{(Figure~\hyperref[fig:ZoneOptimization_Figure_1]{\ref*{fig:ZoneOptimization_Figure_1}.A})} and demonstrate its end-to-end pipeline in Section \ref{subsec:framework_overview}. 
Building on this, we further detail two key components of COMPASS: an encoder producing optimization-ready urban representation in Section \ref{subsec:optimization_ready_urban_representation}, and a difficulty signal for parameter generation in Section \ref{subsec:difficulty_aware_parameter_generation}.

\subsection{Framework Overview}
\label{subsec:framework_overview}

\begin{figure}[t]
\centering
\IfFileExists{figures/ZoneOptimization_Figure_1.pdf}{\includegraphics[width=\textwidth,keepaspectratio]{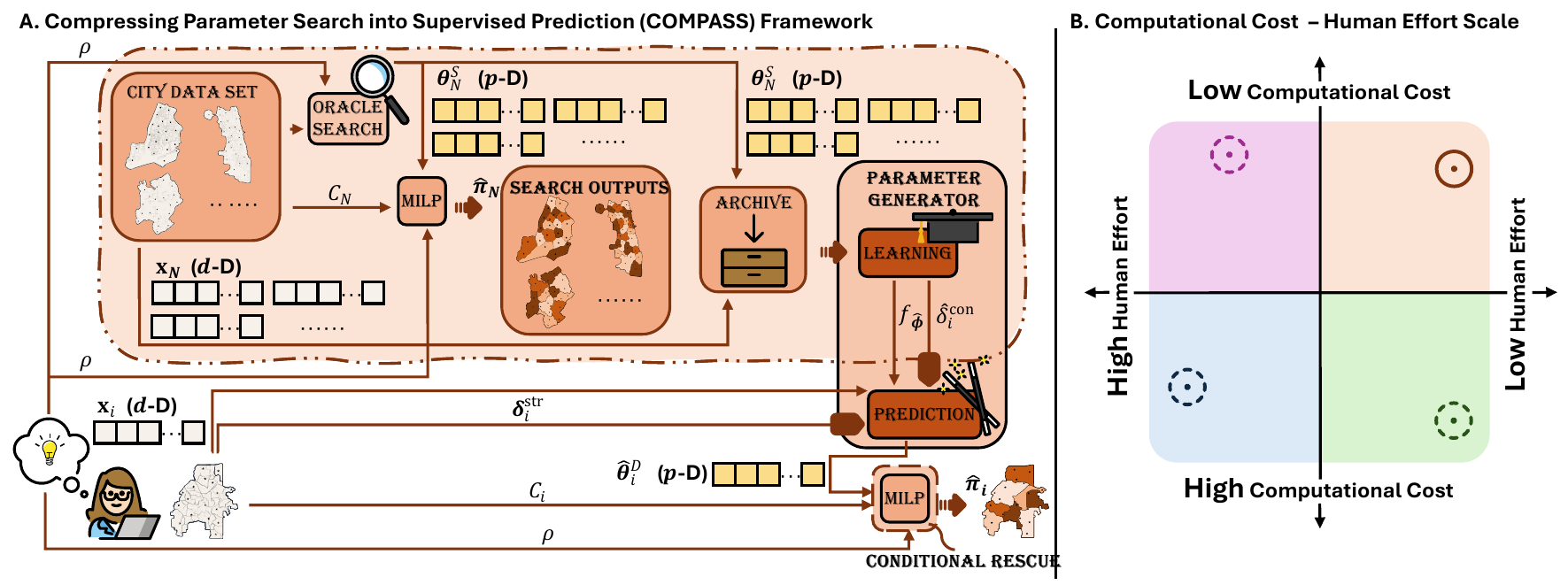}}{\fbox{\parbox[c][3.25cm][c]{0.88\textwidth}{figures/ZoneOptimization_Figure_1.pdf}}}
\caption{(\textit{A}) COMPASS compresses offline parameter search into supervised learning and makes parameter prediction for an unseen city. (\textit{B}) COMPASS and the three conventional strategies are compared in term of city-specific computational cost and human effort.}
\label{fig:ZoneOptimization_Figure_1}
\end{figure}

To simultaneously reduce computational cost and human effort in school zoning policy transfer, we propose COMPASS, with two main stages: Offline Learning and Target-City Deployment.
Through these stages, COMPASS leverages historical search experience to obtain school zoning plan for any unseen city through a single parameter-generation step.
The end-to-end pipeline of COMPASS and its effectiveness \textbf{(Figure~\hyperref[fig:ZoneOptimization_Figure_1]{\ref*{fig:ZoneOptimization_Figure_1}.B})} are presented in Figure \ref{fig:ZoneOptimization_Figure_1}.

\paragraph{Offline Learning.}
Two modules are active in the Offline Learning stage: the Oracle Search and the Parameter Generator.
For each training city, we first run an Oracle Search on its complete city instance \(C_i\) under the fixed policy specification \(\boldsymbol{\rho}\), which filters out a sequence of candidate control vectors.
Each candidate is passed to the downstream MILP to examine whether it can produce a feasible school zoning plan.
Among the feasible candidates, we compare them from a policy perspective and define the best one as the teacher control vector \(\boldsymbol{\theta}_i^S\).
We now pair each training city's optimization-ready urban representation \(\mathbf{x}_i=\mathcal{E}(C_i)\) with its teacher control vector \(\boldsymbol{\theta}_i^S\) to form a cross-city supervision archive: \(\mathcal D^S=\left\{\left(\mathbf x_i,\boldsymbol{\theta}_i^S\right)\right\}_{i=1}^{N}\).
Finally, we use this archive to train the Parameter Generator, thereby establishing a mapping from city context to interpretable criteria for school zoning.
At this point, COMPASS is ready for deployment on a target city without re-running the offline stage.

\paragraph{Target-City Deployment.}
The Target-City Deployment stage involves three consecutive modules: the Parameter Generator, the MILP, and the Conditional Rescue.
To deploy COMPASS to an unseen target city, we must encode the complete city instance \(C_{\mathrm{new}}\) into an optimization-ready urban representation \(\mathbf{x}_{\mathrm{new}}=\mathcal E(C_{\mathrm{new}})\) in advance.
Once the representation is ready, the trained Parameter Generator directly predicts a city-specific control vector, based on its urban representation: \(\overline{\boldsymbol{\theta}}_{\mathrm{new}}=g_{\widehat{\boldsymbol{\phi}}}\left(\mathbf{x}_{\mathrm{new}}\right)\in\Theta\).
Next, we combine this control vector \(\overline{\boldsymbol{\theta}}_{\mathrm{new}}\) with the complete city instance of the target city \(C_{\mathrm{new}}\) to initialize the MILP configuration \(\boldsymbol{\omega}_{\mathrm{new}}^{[0]}=\mathcal R\!\left(C_{\mathrm{new}},\overline{\boldsymbol{\theta}}_{\mathrm{new}}\right)\).
The downstream MILP then utilizes this configuration under the policy specification \(\boldsymbol{\rho}\), generating an explicit zoning plan \(
\widehat{\boldsymbol{\pi}}_{\mathrm{new}}^{[0]}=\operatorname{MILP}\!\left(C_{\mathrm{new}};\boldsymbol{\omega}_{\mathrm{new}}^{[0]},\boldsymbol{\rho}\right)\).
Typically, the MILP would directly generate a feasible solution; otherwise, if the initial MILP attempt is infeasible or hits the time limit, a Conditional Rescue is triggered.
It repairs the control vector in a predefined order, re-solves the MILP, and returns the first successful zoning plan.

\subsection{Optimization-ready Urban Representation}
\label{subsec:optimization_ready_urban_representation}

To allow the same Parameter Generator to handle heterogeneous cities, we adopt a shared encoder \(\mathcal E\) to encode both training cities and unseen target cities into optimization-ready urban representations.
The encoder \(\mathcal E\) summarizes each complete city instance into a set of city-level numerical indicators, which influence the feasibility and policy outcomes of school zoning plans.
Specifically, these indicators can be naturally divided into five feature groups: internal spatial structure \(\mathbf{x}_i^{\mathrm{spatial}}\), school supply \(\mathbf{x}_i^{\mathrm{school}}\), student demand \(\mathbf{x}_i^{\mathrm{student}}\), demographic composition and equity conditions \(\mathbf{x}_i^{\mathrm{equity}}\), and the optimization context \(\mathbf{x}_i^{\mathrm{context}}\).
Table \ref{tab:urban_representation} presents the blocks, symbols, counts, and definitions of each group.
They collectively form a \(d\)-dimensional optimization-ready urban representation, where \(d=140\): 
\begin{equation}
\label{eq:140_dimensional_urban_presentation}
\mathbf{x}_i=\left[\mathbf{x}_i^{\mathrm{spatial}},\mathbf{x}_i^{\mathrm{school}},\mathbf{x}_i^{\mathrm{student}},\mathbf{x}_i^{\mathrm{equity}},\mathbf{x}_i^{\mathrm{context}}\right]\in\mathbb R^{d}.
\end{equation}
This resulting representation therefore serves as a common input interface for COMPASS.

\begin{table}[t]
\centering
\caption{Composition of the 140-dimensional optimization-ready urban representation.}
\label{tab:urban_representation}
\begin{tabular*}{\textwidth}{
@{\extracolsep{\fill}}llcl@{}}
\toprule
Block & Symbol & Count & Definition \\
\midrule
Spatial structure & \(\mathbf{x}_i^{\mathrm{spatial}}\) & 18 & Scale, connectivity, and internal organization\\
School supply & \(\mathbf{x}_i^{\mathrm{school}}\) & 39 & Availability, capacity, distribution and quality of schools\\
Student demand & \(\mathbf{x}_i^{\mathrm{student}}\) & 10 & Volume and distribution of students across spatial units\\
Demographic equity & \(\mathbf{x}_i^{\mathrm{equity}}\) & 27 & Population heterogeneity and policy-relevant groups' distribution\\
Optimization context & \(\mathbf{x}_i^{\mathrm{context}}\) & 46 & Policy preferences, parameter domains, and evaluation settings\\
\bottomrule
\end{tabular*}
\end{table}

\subsection{Difficulty-aware Parameter Generation}
\label{subsec:difficulty_aware_parameter_generation}

Cities vary substantially in the difficulty of satisfying constraints when generating feasible school zoning plans.
Therefore, we must take variation in zoning difficulty across cities into consideration.
Accordingly, COMPASS deliberately incorporates difficulty at three points: difficulty-signal construction, difficulty-sensitive asymmetric learning, and difficulty-dependent parameter relaxation.

\paragraph{Difficulty-signal construction.}
To make the Parameter Generator in our COMPASS aware of difficulty differences, we first construct difficulty signals.
The complete difficulty signals \(\boldsymbol{\delta}_i\) consist of two kinds of complementary information: a structural difficulty profile \(\boldsymbol{\delta}_i^{\mathrm{str}}\) and a plan-generation difficulty label \(\delta_i^{\mathrm{gen}}\).
The structural difficulty profile describes the potential difficulty arising from the city's own conditions, including spatial-service difficulty, capacity tightness, school-distribution imbalance, demographic-constraint risk, and graph-fragmentation risk:
\begin{equation}
\label{eq:structural_difficulty_profile}
{\boldsymbol{\delta}}_i^{\mathrm{str}}=\left[{\delta}_i^{\mathrm{spatial}},{\delta}_i^{\mathrm{capacity}},{\delta}_i^{\mathrm{school}},{\delta}_i^{\mathrm{demographic}},{\delta}_i^{\mathrm{graph}}\right].
\end{equation}
Table \ref{tab:structural_profile} presents the dimensions, symbols, levels, and zoning effects of each difficulty.
We calculate structural difficulty of both training cities and unseen target cities directly from their own urban representations.
In contrast, the plan-generation difficulty label summarizes the overall execution difficulty during the generation of a feasible school zoning plan.
For training cities, this label is derived from the actual search records of the Oracle Search and the corresponding downstream MILP; while for unseen target cities, COMPASS trains an auxiliary classifier to predict the level of plan-generation difficulty:
\begin{equation}
\label{eq:plan_generation_difficulty_label}
\widehat{\delta}_i^{\mathrm{gen}}=h_{\widehat{\boldsymbol{\psi}}}\left(\mathbf{x}_i\right)\in\{\mathrm{Easy},\mathrm{Moderate},\mathrm{Hard},\mathrm{Extreme}\}.
\end{equation}

\begin{table}[t]
\centering
\caption{Composition of the structural difficulty profile.}
\label{tab:structural_profile}
\begin{tabular*}{\textwidth}{
@{\extracolsep{\fill}}llcl@{}}
\toprule
Difficulty dimension & Symbol & Levels & Zoning effect \\
\midrule
Spatial-service difficulty & \({\delta}_i^{\mathrm{spatial}}\) & Low / High & Service distance and spatial coverage \\
Capacity tightness & \({\delta}_i^{\mathrm{capacity}}\) & Loose / Tight & Capacity balance and shortage feasibility \\
School-distribution imbalance & \({\delta}_i^{\mathrm{school}}\) & Low / High & School balance and zoning granularity \\
Demographic-constraint risk & \({\delta}_i^{\mathrm{demographic}}\) & Low / High & Simultaneous satisfaction of equity bounds \\
Graph-fragmentation risk & \({\delta}_i^{\mathrm{graph}}\) & Low / High & Contiguity and spatial feasibility \\
\bottomrule
\end{tabular*}
\end{table}

\paragraph{Difficulty-sensitive asymmetric learning.}
Besides using difficulty signals as input, we also make the training objective of the Parameter Generator difficulty-sensitive.
Such a training objective is necessary because the prediction errors of the control vectors affect the downstream MILP differently depending on the deviation direction.
That is, under the same error magnitude, when the error shifts to a feasibility-tightening direction, the school-zoning problem can be infeasible, whereas an error in a relatively safe direction is less likely to pose such a risk.
Facing this asymmetry, COMPASS penalizes the prediction errors in the risky direction more heavily, using an asymmetric loss:
\begin{equation}
\label{eq:asymmetric_loss}
\mathcal{L}_{\mathrm{asym}}=\frac{1}{Np}\sum_{i=1}^N\sum_{j=1}^p w_{ij}e_{ij}^2,
\end{equation}
where \(e_{ij}=\widehat{\theta}_{ij}^{D}-\theta_{ij}^{S}\) refers to the difference between the raw predicted value and the teacher value of control parameter \(j\) for city \(i\), \(w_{ij}\) indicates the difficulty-sensitive weight, \(N\) and \(p\) denote the number of training cities and the dimension of the control vector, respectively.
By minimizing this loss, the Parameter Generator finishes its training.

\paragraph{Difficulty-dependent parameter relaxation.}
Although asymmetric learning increases the penalty on errors in the risky direction, it still can not rule out the case that several moderate errors drift simultaneously toward the tightening direction, making the downstream MILP infeasible.
To resolve this case, we perform a parameter relaxation before constructing the MILP configuration, which adjusts the control vector depending on the plan-generation difficulty label \(\widehat{\delta}_i^{\mathrm{gen}}\).
The higher the predicted difficulty level, the more extensive the adjustment applied to the control vector.
After the difficulty-dependent parameter relaxation, COMPASS processes the adjusted control vector following the remaining Target-City Deployment procedure introduced in Section \ref{subsec:framework_overview}, to generate an explicit school-zoning plan.

\section{Results}
\label{sec:results}

In this section, we empirically validate the effectiveness of COMPASS, our solution for \hyperref[rq:q1]{\textbf{Q1}}.
We first evaluate the aggregate performance of COMPASS across all test cities in Section \ref{subsec:experimental_evaluation}.
And then, we give
an in-detailed analysis to demonstrate the exact zoning effect of COMPASS in Section \ref{subsec:case_study_atlanta}.

\subsection{Experimental Evaluation}
\label{subsec:experimental_evaluation}

To validate COMPASS at the aggregate level, we set up our experimental evaluation based on a self-constructed school zoning dataset spanning 3,000 US cities.
We adopt 5-fold cross-validation, splitting the cities into training and test set at a ratio of 80:20 in each fold.
After finishing the fold validations, each city is evaluated exactly once, thus yielding out-of-fold results across all 3,000 cities.
Based on these results, we examine COMPASS from three aspects: parameter prediction, policy recovery, and computational efficiency.

\paragraph{Parameter prediction.}
We first test the learning capability of COMPASS, namely, its prediction accuracy on the parameters of the control vector.
To quantify this prediction accuracy, we compare the predicted control parameters with the search-derived values and compute the standard regression metrics.
In general, COMPASS achieves a macro-averaged MAE of 0.539, a macro-averaged RMSE of 0.986, and a macro-averaged \(R^2\) of 0.360 on the out-of-fold predictions across all 3,000 cities.
These results indicate that the Parameter Generator can learn a reliable parameter-prediction model from the supervision provided by the Oracle Search.
Thus, COMPASS can capture the cross-city patterns linking urban representations and school-zoning control vectors. 

\begin{figure}[t]
\centering
\includegraphics[width=0.82\textwidth,keepaspectratio]{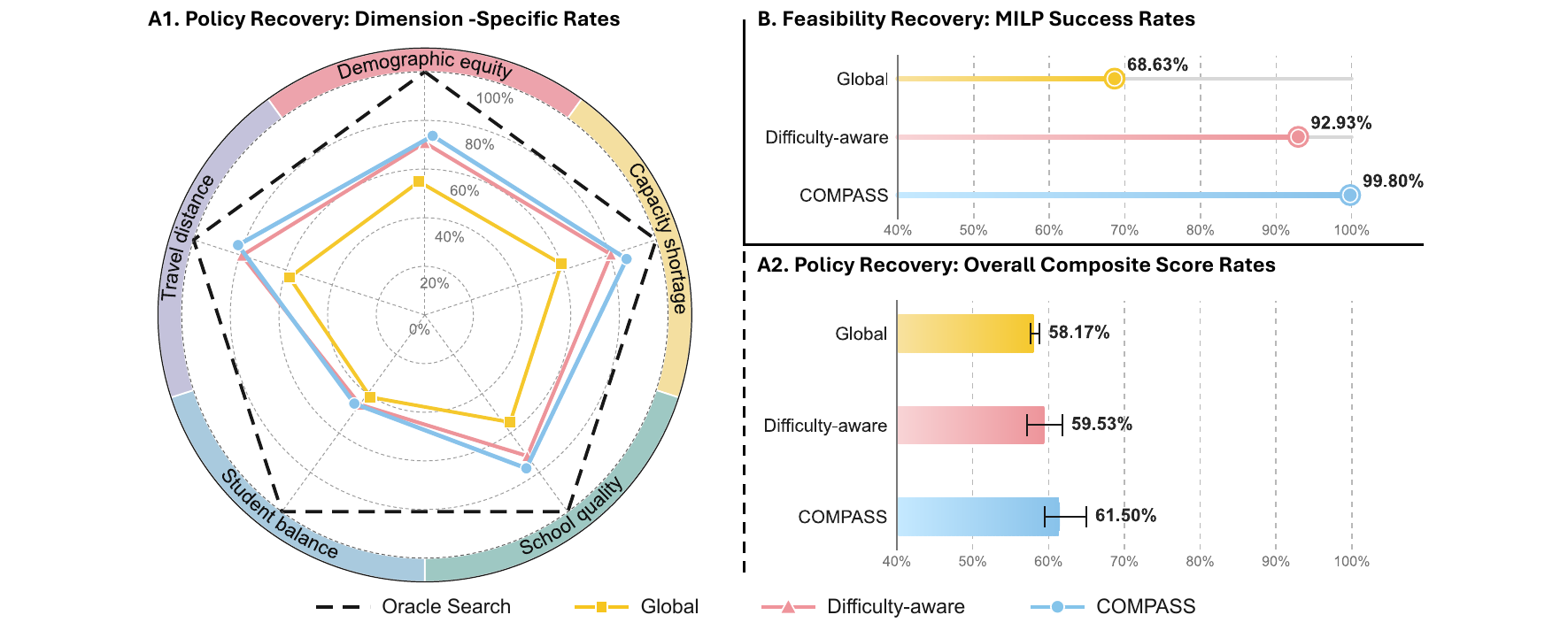}
\caption{(\textit{A}) End-to-end policy recovery of COMPASS and its two ablation variants over all held-out cities: (\textit{A1}) within-tolerance recovery rates across five policy dimensions, and (\textit{A2}) recovery rates for the composite policy score, with whiskers showing the observed range across the 5 cross-validation folds. (\textit{B}) Solver-feasible MILP success rates.}
\label{fig:ZoneOptimization_Figure_2}
\end{figure}

\paragraph{Policy recovery.}
Since the parameter-prediction accuracy does not guarantee the feasibility of the corresponding school zoning plan, we further evaluate COMPASS's policy recovery performance.
A city is recorded as recovered only if the following two conditions are simultaneously met: 
\begin{enumerate}
    \item Its predicted control vector enables the downstream MILP to successfully return a feasible zoning plan;
    \item The actual policy outcome of this plan falls within a preset tolerance of the search-derived outcome.
\end{enumerate}
For COMPASS and its two ablation variants, the solver-feasible success rate of the downstream MILP and the within-tolerance recovery rates across all policy dimensions are reported in Figure \ref{fig:ZoneOptimization_Figure_2}.
These progressive improvements in the ablation results demonstrate that both the Difficulty-aware Parameter Generation and the Conditional Rescue additionally contribute to the policy recovery.
These contributions make COMPASS reach a solver-feasible success rate of 99.80\% \textbf{(Figure~\hyperref[fig:ZoneOptimization_Figure_2]{\ref*{fig:ZoneOptimization_Figure_2}.B})}, with all reported policy dimensions achieving the highest recovery rates \textbf{(Figure~\hyperref[fig:ZoneOptimization_Figure_2]{\ref*{fig:ZoneOptimization_Figure_2}.A1})}.
Overall, COMPASS generates feasible school zoning plans for new cities while effectively recovering the key policy outcomes \textbf{(Figure~\hyperref[fig:ZoneOptimization_Figure_2]{\ref*{fig:ZoneOptimization_Figure_2}.A2})}.

\begin{figure}[t]
\centering
\includegraphics[width=\textwidth,keepaspectratio]{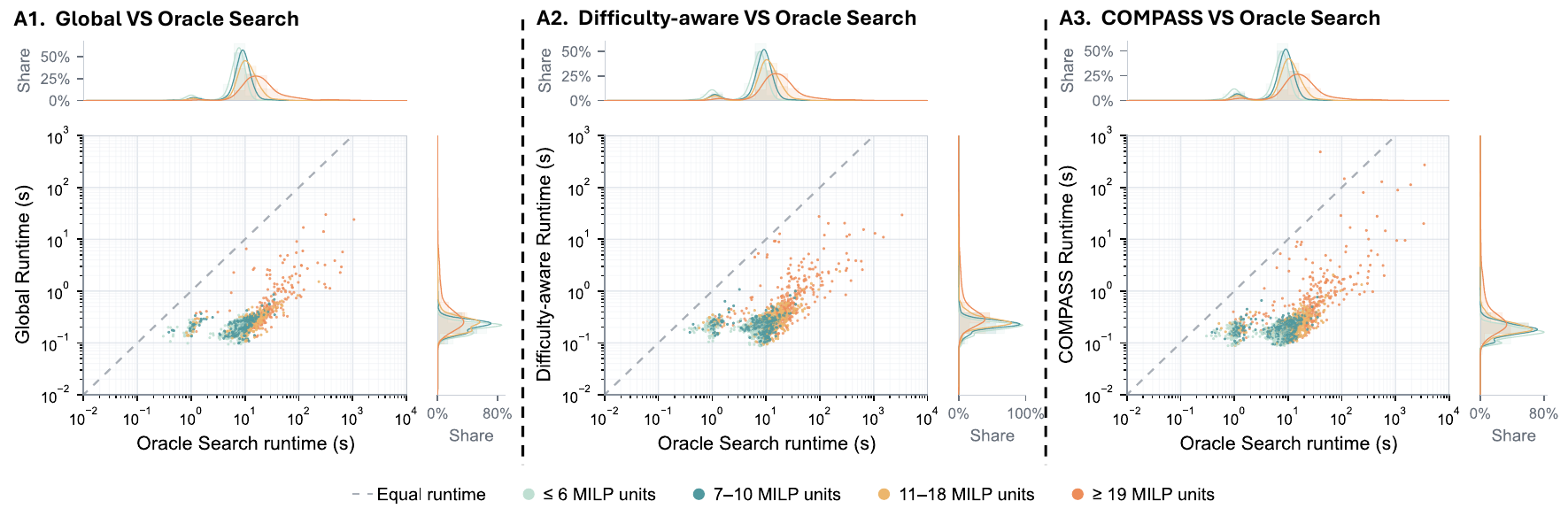}
\caption{(\textit{A}) Paired city-level runtime comparisons of COMPASS and its two ablation variants against Oracle Search across four city-scale strata defined by assignment-unit count: (\textit{A1}) Global versus Oracle Search, (\textit{A2}) Difficulty-aware versus Oracle Search, and (\textit{A3}) COMPASS versus Oracle Search, with the dashed line indicating equal runtime and the marginal distributions summarizing runtime variation under the paired approaches.}
\label{fig:ZoneOptimization_Figure_4}
\end{figure}

\paragraph{Computational efficiency.}
Finally, we evaluate COMPASS's computational efficiency.
Specifically, we measure the runtime for new cities with a feasible school zoning plan.
According to this measurement, Table \ref{tab:runtime_performance} summarizes the deployment-stage runtime of different methods.
Compared with repeating Oracle Search for each new target city, COMPASS achieves a 49.14\(\times\) median runtime speedup.
We further compare the runtime performance of COMPASS and its two ablation variants \textbf{(Figure~\hyperref[fig:ZoneOptimization_Figure_4]{\ref*{fig:ZoneOptimization_Figure_4}.A1} \& \hyperref[fig:ZoneOptimization_Figure_4]{\ref*{fig:ZoneOptimization_Figure_4}.A2})}, finding that the efficiency advantage of COMPASS is maintained across all city scales \textbf{(Figure~\hyperref[fig:ZoneOptimization_Figure_4]{\ref*{fig:ZoneOptimization_Figure_4}.A3})}.
Taken together, COMPASS significantly reduces the computational cost and improves the deployment efficiency for unseen cities.

\begin{table}[t]
\centering
\caption{Runtime comparison between Oracle Search and \COMPASSicon COMPASS on successful runs.}
\label{tab:runtime_performance}
\begin{tabular*}{\textwidth}{@{\extracolsep{\fill}}lcccc@{}}
\toprule
Method & Median runtime (s) & IQR (s) & Median Speedup (vs. Oracle Search) \\
\midrule
Oracle Search & 9.150 & 7.059 - 12.905  & 1\(\times\) \\
\COMPASSicon COMPASS & \textbf{0.186} & \textbf{0.155 - 0.236} & \textbf{49.14\(\times\)} \\
\bottomrule
\end{tabular*}
\end{table}

\subsection{Case Study: Atlanta}
\label{subsec:case_study_atlanta}

\begin{figure}[t]
\centering
\includegraphics[width=\textwidth,keepaspectratio]{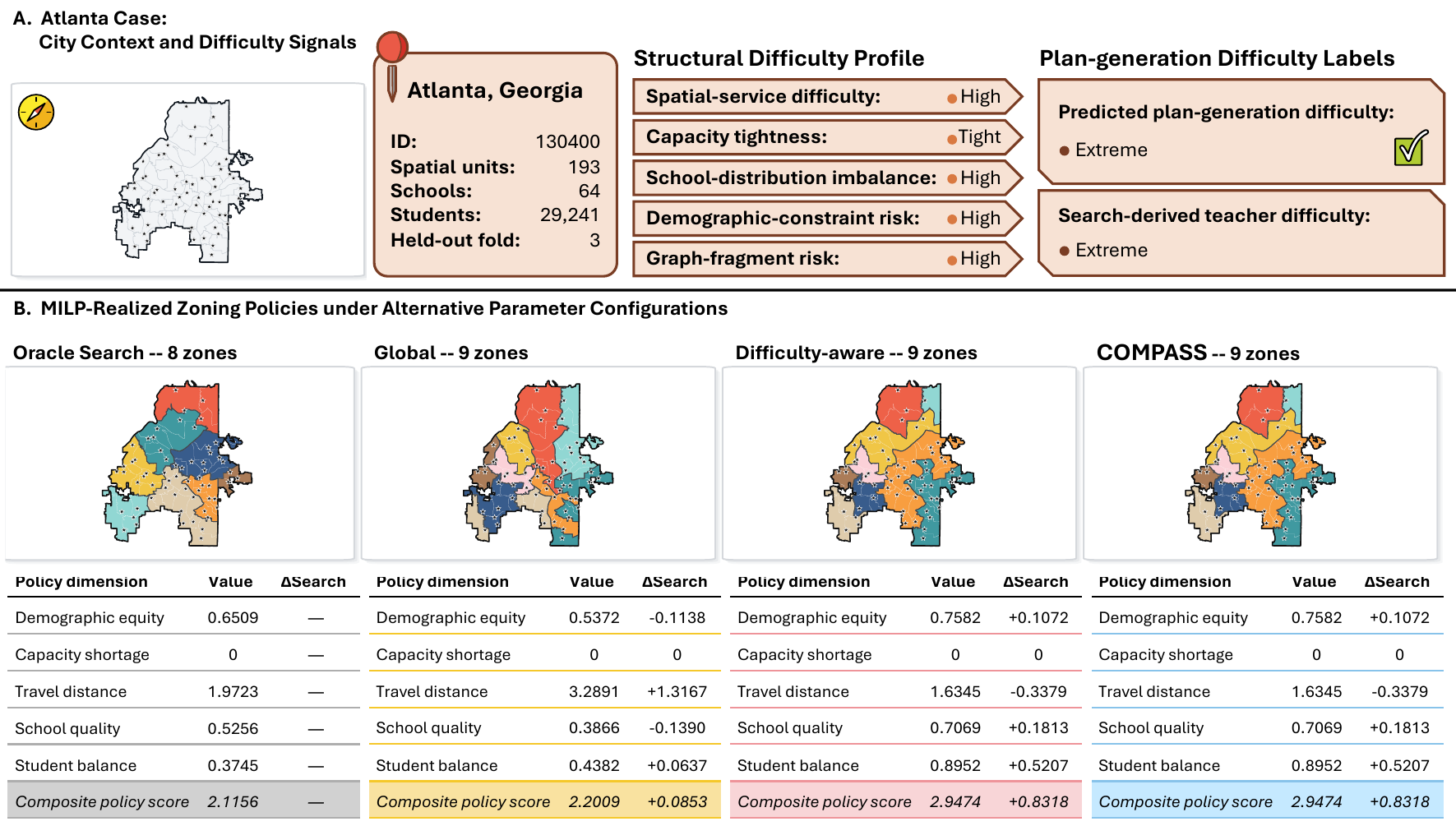}
\caption{(\textit{A}) A map presents the city context of Atlanta together with its structural difficulty profile and plan-generation difficulty labels. (\textit{B}) Zoning plans and corresponding policy outcomes generated by COMPASS are compared with its two ablation variants and the search-derived reference.}
\label{fig:ZoneOptimization_Figure_5}
\end{figure}

To evaluate the performance of COMPASS on a concrete unseen city, we randomly select a test city from fold 3, Atlanta, Georgia, as a case study \textbf{(Figure~\hyperref[fig:ZoneOptimization_Figure_5]{\ref*{fig:ZoneOptimization_Figure_5}.A})}.
Atlanta comprises 193 spatial units, 64 schools, and 29,241 students.
For this case, we adopt an experimental policy specification, which prioritizes demographic equity, followed by capacity adequacy, school-quality balance, travel distance, and so on.
Under the same Atlanta city context and this fixed policy preference, we compare COMPASS with its two ablation variants and the search-derived reference in Figure \ref{fig:ZoneOptimization_Figure_5}.

Among all detectable results, we first observe that the auxiliary classifier correctly predicts Atlanta's plan-generation difficulty as Extreme.
Beyond this component-level result, we find that COMPASS successfully generates a feasible zoning plan for Atlanta \textbf{(Figure~\hyperref[fig:ZoneOptimization_Figure_5]{\ref*{fig:ZoneOptimization_Figure_5}.B})}.
Notably, we find that the difficulty-aware configuration is enough to generate feasible zoning plans, even without triggering the Conditional Rescue, the zoning plan of COMPASS is identical to its difficulty-aware ablation variant.
This plan ensures zone contiguity, maintains 0 capacity shortage, and reduces travel distance by 17.13\%, which also manages the trade-off across all policy dimensions through a carefully designed compromise.
This city-level evidence shows that COMPASS can support the transferability of school attendance zoning in practice.

\section{Discussion}
\label{sec:discussion}

We answer the empirical question \hyperref[rq:q2]{\textbf{Q2}} in Section \ref{subsec:transferability_patterns} by analyzing the quantitative evidence reported in Section \ref{sec:results} and summarizing the transferability patterns therein.
We also discuss the boundary of responsibility for local decision-makers when deploying COMPASS and then carrying its outputs forward into subsequent policy implementation in Section \ref{subsec:policy_implementation}.

\subsection{Transferability Patterns}
\label{subsec:transferability_patterns}

\paragraph{Urban contexts.}
The transferability benefits brought by COMPASS are not evenly distributed across urban contexts.
Although COMPASS performs well in general, it mainly benefits cities that originally struggled to obtain feasible school zoning plans.
These difficult cities are largely constrained by the structure of urban resources.
For example, when multiple adverse conditions appear simultaneously, such as tight capacity and imbalanced school distribution, the risk of infeasible parameter configurations increases significantly.
In contrast, city scale itself does not directly determine zoning difficulty.
A larger city with more spatial units, schools, and students only leads to a higher computational burden and a longer solving time, but the failure rate is not necessarily higher, and obtaining a feasible plan is not necessarily more difficult.

\paragraph{Policy dimensions.}
The transferability varies significantly across different policy dimensions.
In COMPASS, student balance is consistently the hardest to recover, while capacity shortage remains the least difficult, and demographic equity, school quality, and travel distance fall in between.
The differences are influenced by changes in policy preferences to some degree, but this does not imply that the heavier the weight assigned to a policy dimension, the easier it is to transfer.
That is because the final transfer outcomes are jointly determined by urban contexts, the complete policy objectives, and downstream algorithms; thus, they can not be attributed to any single policy weight.

\subsection{Policy Implementation}
\label{subsec:policy_implementation}

For local decision-makers, COMPASS provides two complementary outputs: a MILP-feasible explicit zoning plan and a corresponding policy-interpretable control vector.
The zoning plan allows local decision-makers to see the spatial assignment and the trade-offs across policy dimensions, while the control vector makes the control logic behind the plan inspectable, interpretable, and adjustable to local decision-makers.
Hence, they can add or modify the local constraints, re-weight the policy dimensions, and re-run the downstream MILP to generate zoning plans based on their policy preference.
After comparing multiple feasible plans, they choose the most satisfactory one and formally review it.
During the review, local decision-makers should enforce local laws and regulations, consult community residents and gather their feedback, and ultimately implement the adopted plan.
After the policy implementation, local decision-makers still need to keep monitoring its practical effects.
These responsibilities must be assumed by decision-makers rather than by COMPASS.

\section{Conclusion}
\label{sec:conclusion}

This work enables transferable school attendance zoning by integrating optimization-ready urban representation and difficulty-aware parameter generation into the parameter-to-policy process of COMPASS.
This combination, verified in both aggregate evaluation and individual case study, contributes to control-vector learning, promotes policy recovery, and improves computational efficiency.
These differences in performance further suggest that the transferability of school attendance zoning is systematically shaped by the urban contexts and the policy dimensions.

\bibliographystyle{unsrt}
\bibliography{bib.bib}

\clearpage
\begingroup
\setcounter{section}{0}
\setcounter{subsection}{0}
\setcounter{table}{0}
\setcounter{figure}{0}
\setcounter{equation}{0}
\setcounter{algorithm}{0}
\setcounter{theorem}{0}
\renewcommand{\thesection}{S\arabic{section}}
\renewcommand{\thetable}{S\arabic{table}}
\renewcommand{\thefigure}{S\arabic{figure}}
\renewcommand{\theequation}{S\arabic{equation}}
\renewcommand{\thealgorithm}{S\arabic{algorithm}}
\renewcommand{\thetheorem}{S\arabic{theorem}}
\renewcommand{\theHsection}{supplement.\arabic{section}}
\pdfbookmark[0]{Supplementary Information}{supplementary-information}
\begin{center}
    {\Large\bfseries Supplementary Information\par}
\end{center}

\section{Cross-validation and Runtime Robustness}
\label{sec:fold_wise_cross_validation_results}

To assess whether the aggregate findings are robust to the held-out split, we report pooled and fold-wise results for raw parameter prediction, solver-feasible MILP success, end-to-end within-tolerance policy recovery, policy quality among successfully solved COMPASS cities, and computational efficiency. We additionally report runtime performance across city-scale strata. These disaggregated results verify that the main-text findings are not driven by a single cross-validation fold or a narrow range of city sizes.

\begin{table}[h!t]
\centering
\caption{Aggregate and fold-wise raw out-of-fold parameter-prediction performance of the Global and Difficulty-aware ablation variants.}
\label{tab:parameter_prediction_5fold}
\begin{tabular*}{\textwidth}{
@{\extracolsep{\fill}}ccccccc@{}}
\toprule
\multirow{2}{*}{Fold} & \multicolumn{3}{c}{Global} & \multicolumn{3}{c}{Difficulty-aware} \\
\cmidrule(lr){2-4}\cmidrule(lr){5-7} 
& MAE & RMSE & \(R^2\) & MAE & RMSE & \(R^2\) \\
\midrule
1 & 0.541296 & 1.051589 & 0.577188 & 0.565192 & 1.125570 & 0.394887 \\
2 & 0.529261 & 0.762202 & 0.562072 & 0.507466 & 0.879769 & 0.370511 \\
3 & 0.497018 & 0.712731 & 0.671131 & 0.519224 & 0.917314 & 0.415814 \\
4 & 0.541350 & 0.777686 & 0.528308 & 0.565432 & 1.034077 & 0.187125 \\
5 & 0.538139 & 0.806396 & 0.501145 & 0.537235 & 0.943834 & 0.276450 \\
\midrule
\textit{All}
& \textit{0.529413}
& \textit{0.834521}
& \textit{0.578509}
& \textit{0.538910}
& \textit{0.985823}
& \textit{0.360189} \\
\bottomrule
\end{tabular*}
\end{table}

Across all out-of-fold predictions, the per-parameter \(R^2\) values ranged from 0.433 to 0.784 for the Global ablation and from 0.085 to 0.769 for the raw Difficulty-aware ablation. These metrics are computed before difficulty-dependent parameter relaxation, validity projection, and city-specific realization.

\begin{table}[h!t]
\centering
\caption{Aggregate and fold-wise solver-feasible MILP performance of COMPASS and its two ablation variants.}
\label{tab:feasibility_performance_5fold}
\small
\begin{tabular*}{\textwidth}{
@{\extracolsep{\fill}}llcccccc@{}}
\toprule
Method & Outcome & Fold 1 & Fold 2 & Fold 3 & Fold 4 & Fold 5 & Overall \\
\midrule
\multirow{4}{*}{Global}
& Successful & 414 & 422 & 403 & 423 & 397 & 2,059 \\
& Infeasible & 186 & 176 & 194 & 173 & 200 & 929 \\
& Timeout & 0 & 2 & 3 & 4 & 3 & 12 \\
& \textit{Success rate}
& \textit{69.00\%} & \textit{70.33\%} & \textit{67.17\%}
& \textit{70.50\%} & \textit{66.17\%} & \textit{68.63\%} \\
\midrule
\multirow{4}{*}{Difficulty-aware}
& Successful & 560 & 563 & 559 & 561 & 545 & 2,788 \\
& Infeasible & 40 & 35 & 38 & 36 & 51 & 200 \\
& Timeout & 0 & 2 & 3 & 3 & 4 & 12 \\
& \textit{Success rate}
& \textit{93.33\%} & \textit{93.83\%} & \textit{93.17\%}
& \textit{93.50\%} & \textit{90.83\%} & \textit{92.93\%} \\
\midrule
\multirow{4}{*}{\COMPASSicon COMPASS}
& Successful & 599 & 599 & 598 & 600 & 598 & 2,994 \\
& Infeasible & 1 & 0 & 0 & 0 & 1 & 2 \\
& Timeout & 0 & 1 & 2 & 0 & 1 & 4 \\
& \textit{Success rate}
& \textit{99.83\%} & \textit{99.83\%} & \textit{99.67\%}
& \textit{100.00\%} & \textit{99.67\%} & \textit{99.80\%} \\
\bottomrule
\end{tabular*}
\end{table}

\begin{table}[h!t]
\centering
\caption{Aggregate and fold-wise end-to-end within-tolerance policy recovery relative to Oracle Search.}
\label{tab:policy_recovery_5fold}
\small
\begin{tabular*}{\textwidth}{
@{\extracolsep{\fill}}llcccccc@{}}
\toprule
Method & Policy dimension & Fold 1 & Fold 2 & Fold 3 & Fold 4 & Fold 5 & Overall \\
\midrule
\multirow{6}{*}{Global}
& Demographic equity & 54.83\% & 54.67\% & 53.17\% & 56.83\% & 54.83\% & 54.87\% \\
& Capacity shortage & 59.67\% & 60.33\% & 59.83\% & 60.83\% & 58.83\% & 59.90\% \\
& Travel distance & 57.83\% & 58.00\% & 56.83\% & 60.33\% & 56.00\% & 57.80\% \\
& School quality & 56.50\% & 56.00\% & 57.17\% & 57.00\% & 55.17\% & 56.37\% \\
& Student balance & 39.33\% & 42.00\% & 41.00\% & 40.33\% & 40.00\% & 40.53\% \\
& \textit{Composite policy score}
& \textit{57.83\%} & \textit{58.67\%} & \textit{57.83\%}
& \textit{58.83\%} & \textit{57.67\%} & \textit{58.17\%} \\
\midrule
\multirow{6}{*}{Difficulty-aware}
& Demographic equity & 71.50\% & 70.33\% & 71.50\% & 70.17\% & 70.50\% & 70.80\% \\
& Capacity shortage & 80.17\% & 79.67\% & 83.00\% & 79.83\% & 79.83\% & 80.50\% \\
& Travel distance & 77.50\% & 79.83\% & 79.67\% & 81.67\% & 76.83\% & 79.10\% \\
& School quality & 71.17\% & 70.33\% & 73.83\% & 71.50\% & 71.50\% & 71.67\% \\
& Student balance & 43.67\% & 45.17\% & 47.67\% & 45.00\% & 46.17\% & 45.53\% \\
& \textit{Composite policy score}
& \textit{57.17\%} & \textit{59.33\%} & \textit{59.83\%}
& \textit{59.50\%} & \textit{61.83\%} & \textit{59.53\%} \\
\midrule
\multirow{6}{*}{\COMPASSicon COMPASS}
& Demographic equity & 73.83\% & 72.67\% & 74.00\% & 72.83\% & 75.33\% & 73.73\% \\
& Capacity shortage & 85.33\% & 83.83\% & 88.50\% & 85.67\% & 87.17\% & 86.10\% \\
& Travel distance & 80.50\% & 81.50\% & 82.33\% & 84.50\% & 81.17\% & 82.00\% \\
& School quality & 74.33\% & 72.83\% & 78.50\% & 75.67\% & 76.83\% & 75.63\% \\
& Student balance & 44.33\% & 45.67\% & 48.50\% & 46.17\% & 48.17\% & 46.57\% \\
& \textit{Composite policy score}
& \textit{59.50\%} & \textit{60.50\%} & \textit{61.33\%}
& \textit{61.17\%} & \textit{65.00\%} & \textit{61.50\%} \\
\bottomrule
\end{tabular*}
\end{table}

\begin{table}[h!t]
\centering
\caption{Fold-wise runtime performance of \COMPASSicon COMPASS and its two ablation variants compared with Oracle Search.}
\label{tab:runtime_performance_5fold}
\begin{tabular*}{\textwidth}{
@{\extracolsep{\fill}}ccccccccc@{}}
\toprule
\multirow{2}{*}{Fold} & \multicolumn{4}{c}{Oracle Search} & \multicolumn{4}{c}{Global} \\
\cmidrule(lr){2-5}\cmidrule(lr){6-9}
& Median (s) & IQR (s) & P95 (s) & Speedup & Median (s) & IQR (s) & P95 (s) & Speedup \\
\midrule
1 & 9.177 & 7.217 - 13.186 & 28.496 & 1\(\times\) & 0.221 & 0.172 - 0.269 & 0.528 & 41.52\(\times\) \\
2 & 9.345 & 7.251 - 13.493 & 27.860 & 1\(\times\) & 0.229 & 0.190 - 0.270 & 0.527 & 40.89\(\times\) \\
3 & 8.939 & 7.060 - 12.394 & 28.533 & 1\(\times\) & 0.226 & 0.187 - 0.268 & 0.441 & 39.55\(\times\) \\
4 & 9.374 & 7.038 - 13.619 & 40.098 & 1\(\times\) & 0.215 & 0.167 - 0.270 & 0.790 & 43.62\(\times\) \\
5 & 8.837 & 6.605 - 12.317 & 34.587 & 1\(\times\) & 0.219 & 0.167 - 0.264 & 0.513 & 40.29\(\times\) \\
\midrule
\textit{ALL}
& \textit{9.150} & \textit{7.059 - 12.905} & \textit{31.943} & \textit{1\(\times\)}
& \textit{0.223} & \textit{0.175 - 0.267} & \textit{0.525} & \textit{41.08\(\times\)} \\
\toprule
\multirow{2}{*}{Fold} & \multicolumn{4}{c}{Difficulty-aware} & \multicolumn{4}{c}{\COMPASSicon COMPASS} \\
\cmidrule(lr){2-5}\cmidrule(lr){6-9}
& Median (s) & IQR (s) & P95 (s) & Speedup & Median (s) & IQR (s) & P95 (s) & Speedup \\
\midrule
1 & 0.246 & 0.218 - 0.299 & 0.682 & 37.32\(\times\) & 0.167 & 0.145 - 0.204 & 0.420 & 55.07\(\times\) \\
2 & 0.245 & 0.218 - 0.294 & 0.581 & 38.20\(\times\) & 0.191 & 0.170 - 0.237 & 0.508 & 48.90\(\times\) \\
3 & 0.226 & 0.196 - 0.280 & 0.641 & 39.54\(\times\) & 0.153 & 0.106 - 0.208 & 0.449 & 58.29\(\times\) \\
4 & 0.235 & 0.188 - 0.289 & 0.999 & 39.81\(\times\) & 0.232 & 0.200 - 0.285 & 0.928 & 40.38\(\times\) \\
5 & 0.229 & 0.191 - 0.279 & 0.596 & 38.54\(\times\) & 0.181 & 0.158 - 0.216 & 0.550 & 48.90\(\times\) \\
\midrule
\textit{All}
& \textit{0.237} & \textit{0.203 - 0.289} & \textit{0.688} & \textit{38.65\(\times\)}
& \textit{0.186} & \textit{0.155 - 0.236} & \textit{0.596} & \textit{49.14\(\times\)} \\
\bottomrule
\end{tabular*}
\end{table}

We further evaluated runtime across four city-scale strata defined by the number of assignment units entering the zoning MILP. Table \ref{tab:runtime_size_strata} reports the median runtime among successful runs within each stratum. The corresponding paired city-level distributions are presented in Figure~\ref{fig:ZoneOptimization_Figure_4} of the main text.

\begin{table}[h!t]
\centering
\caption{Median runtime across city-scale strata defined by assignment-unit count among successful runs.}
\label{tab:runtime_size_strata}
\begin{tabular*}{\textwidth}{
@{\extracolsep{\fill}}lccccc@{}}
\toprule
City-scale stratum & Assignment units & Oracle Search & Global
& Difficulty-aware & \COMPASSicon COMPASS \\
\midrule
Q1 & \(n_i \leq 6\) & 7.179 s & 0.201 s & 0.208 s & 0.159 s \\
Q2 & \(7 \leq n_i \leq 10\) & 8.799 s & 0.220 s & 0.227 s & 0.180 s \\
Q3 & \(11 \leq n_i \leq 18\) & 10.488 s & 0.230 s & 0.249 s & 0.192 s \\
Q4 & \(n_i \geq 19\) & 16.547 s & 0.287 s & 0.321 s & 0.261 s \\
\bottomrule
\end{tabular*}
\end{table}

\begin{table}[h!t]
\centering
\caption{Aggregate policy recovery performance among successfully solved \COMPASSicon COMPASS cities.}
\label{tab:dimension_policy_recovery}
\begingroup
\newcommand{\tightrow}{\\[-4pt]}
\begin{tabular*}{\textwidth}{
@{\extracolsep{\fill}}lcccc@{}}
\toprule
\multirow{2}{*}{Policy dimension}
& \multicolumn{4}{c}{All successful COMPASS cities (\(n=2,994\))} \\
\cmidrule(lr){2-5}
& Median absolute gap & IQR & Absolute recovery & No-material degradation \\
\midrule
Demographic equity & 0.0033 & 0 - 0.0532 & 73.88\% & 95.29\% \tightrow
Capacity shortage & 0 & 0 - 0 & 86.27\% & 89.01\% \tightrow
Travel distance & 0.0733 & 0 - 0.3710 & 82.16\% & 88.48\% \tightrow
School quality & 0 & 0 - 0.0444 & 75.78\% & 88.34\% \tightrow
Student balance & 0.0533 & 0 - 0.3583 & 46.66\% & 54.61\% \tightrow
\textit{Composite policy score}
& \textit{0.0795} & \textit{0 - 0.4406}
& \textit{61.62\%} & \textit{66.17\%} \\
\bottomrule
\end{tabular*}
\endgroup
\end{table}

\begin{table}[h!t]
\centering
\caption{Fold-wise policy recovery performance among successfully solved \COMPASSicon COMPASS cities.}
\label{tab:dimension_policy_recovery_5fold}
\begingroup
\newcommand{\tightrow}{\\[-4pt]}
\begin{tabular*}{\textwidth}{
@{\extracolsep{\fill}}lcccc@{}}
\toprule
\multirow{2}{*}{Policy dimension} & \multicolumn{4}{c}{Fold 1} \\
\cmidrule(lr){2-5}
& Median absolute gap & IQR & Absolute recovery & No-material degradation \\
\midrule
Demographic equity & 0.0046 & 0 - 0.0536 & 73.96\% & 94.16\% \tightrow
Capacity shortage & 0 & 0 - 0 & 85.48\% & 88.48\% \tightrow
Travel distance & 0.0916 & 0 - 0.4222 & 80.63\% & 87.31\% \tightrow
School quality & 0 & 0 - 0.0552 & 74.46\% & 86.48\% \tightrow
Student balance & 0.0637 & 0 - 0.3703 & 44.41\% & 53.26\% \tightrow
\textit{Composite policy score} & \textit{0.0811} & \textit{0 - 0.4442} & \textit{59.60\%} & \textit{63.94\%} \\
\toprule
\multirow{2}{*}{Policy dimension} & \multicolumn{4}{c}{Fold 2} \\
\cmidrule(lr){2-5}
& Median absolute gap & IQR & Absolute recovery & No-material degradation \\
\midrule
Demographic equity & 0.0056 & 0 - 0.0567 & 72.79\% & 95.16\% \tightrow
Capacity shortage & 0 & 0 - 0 & 83.97\% & 86.98\% \tightrow
Travel distance & 0.0959 & 0 - 0.3799 & 81.64\% & 86.98\% \tightrow
School quality & 0 & 0 - 0.0598 & 72.95\% & 87.65\% \tightrow
Student balance & 0.0693 & 0 - 0.3765 & 45.74\% & 53.09\% \tightrow
\textit{Composite policy score} & \textit{0.1022} & \textit{0 - 0.4552} & \textit{60.60\%} & \textit{64.77\%} \\
\toprule
\multirow{2}{*}{Policy dimension} & \multicolumn{4}{c}{Fold 3} \\
\cmidrule(lr){2-5}
& Median absolute gap & IQR & Absolute recovery & No-material degradation \\
\midrule
Demographic equity & 0.0027 & 0 - 0.0518 & 74.25\% & 95.32\% \tightrow
Capacity shortage & 0 & 0 - 0 & 88.80\% & 90.97\% \tightrow
Travel distance & 0.0564 & 0 - 0.3351 & 82.61\% & 89.63\% \tightrow
School quality & 0 & 0 - 0.0227 & 78.76\% & 89.46\% \tightrow
Student balance & 0.0339 & 0 - 0.3352 & 48.66\% & 56.35\% \tightrow
\textit{Composite policy score} & \textit{0.0711} & \textit{0 - 0.4264} & \textit{61.54\%} & \textit{66.05\%} \\
\toprule
\multirow{2}{*}{Policy dimension} & \multicolumn{4}{c}{Fold 4} \\
\cmidrule(lr){2-5}
& Median absolute gap & IQR & Absolute recovery & No-material degradation \\
\midrule
Demographic equity & 0.0027 & 0 - 0.0533 & 72.83\% & 95.33\% \tightrow
Capacity shortage & 0 & 0 - 0 & 85.67\% & 88.33\% \tightrow
Travel distance & 0.0765 & 0 - 0.3370 & 84.50\% & 90.83\% \tightrow
School quality & 0 & 0 - 0.0448 & 75.67\% & 87.50\% \tightrow
Student balance & 0.0590 & 0 - 0.3737 & 46.17\% & 53.50\% \tightrow
\textit{Composite policy score} & \textit{0.0798} & \textit{0 - 0.4738} & \textit{61.17\%} & \textit{65.50\%} \\
\toprule
\multirow{2}{*}{Policy dimension} & \multicolumn{4}{c}{Fold 5} \\
\cmidrule(lr){2-5}
& Median absolute gap & IQR & Absolute recovery & No-material degradation \\
\midrule
Demographic equity & 0.0023 & 0 - 0.0478 & 75.59\% & 96.49\% \tightrow
Capacity shortage & 0 & 0 - 0 & 87.46\% & 90.30\% \tightrow
Travel distance & 0.0481 & 0 - 0.3973 & 81.44\% & 87.63\% \tightrow
School quality & 0 & 0 - 0.0355 & 77.09\% & 90.64\% \tightrow
Student balance & 0.0359 & 0 - 0.3462 & 48.33\% & 56.86\% \tightrow
\textit{Composite policy score} & \textit{0.0583} & \textit{0 - 0.3961} & \textit{65.22\%} & \textit{70.57\%} \\
\bottomrule
\end{tabular*}
\endgroup
\end{table}

The disaggregated results support the aggregate findings. Raw parameter-prediction patterns were consistent across folds, while Difficulty-aware Parameter Generation and Conditional Rescue progressively expanded solver-feasible coverage. COMPASS achieved solver-feasible success rates between 99.67\% and 100.00\% across the five folds and exceeded the Global ablation in fixed-denominator recovery for every policy dimension in every fold, although student balance remained the least recovered dimension. All learned pipelines retained more than a 37\(\times\) median speedup over Oracle Search in every fold. Runtime also remained substantially lower across all city-scale strata: from Q1 to Q4, the COMPASS median increased from 0.159 to 0.261 seconds, compared with 7.179 to 16.547 seconds for Oracle Search. These patterns indicate that the main findings are not attributable to a single held-out fold or to only the smallest city instances.

\section{Sensitivity to Search Methods}
\label{sec:sensitivity_to_search_methods}

Because the main experiments use Simulated Annealing to generate the search-derived supervision, we first examined whether the performance of COMPASS depends on this particular teacher. We replaced Simulated Annealing with Bayesian Optimization and retained the same Parameter Generator design while using the resulting teacher control vectors, together with the same COMPASS framework and downstream evaluation criteria. Parameter prediction, solver-feasible MILP performance and computational efficiency, and fixed-denominator policy recovery are reported in Tables \ref{tab:search_parameter_prediciton}, \ref{tab:search_feasibility_runtime}, and \ref{tab:search_compass_policy_recovery}, respectively.

\begin{table}[h!t]
\centering
\caption{Parameter prediction performance under alternative search methods.}
\label{tab:search_parameter_prediciton}
\begin{tabular*}{\textwidth}{
@{\extracolsep{\fill}}lcccccc@{}}
\toprule
\multirow{2}{*}{Search method} & \multicolumn{3}{c}{Global} & \multicolumn{3}{c}{Difficulty-aware} \\
\cmidrule(lr){2-4}\cmidrule(lr){5-7} 
& MAE & RMSE & \(R^2\) & MAE & RMSE & \(R^2\) \\
\midrule
Simulated Annealing & 0.529413 & 0.834521 & 0.578509 & 0.538910 & 0.985823 & 0.360189 \\
Bayesian Optimization & 0.529282 & 1.285510 & 0.420286 & 0.776004 & 1.526463 & 0.237476 \\
\bottomrule
\end{tabular*}
\end{table}

\begin{table}[h!t]
\centering
\caption{Feasibility and runtime performance under alternative search methods.}
\label{tab:search_feasibility_runtime}
\begin{tabular*}{\textwidth}{
@{\extracolsep{\fill}}lccccc@{}}
\toprule
\multirow{2}{*}{Search method} & \multicolumn{2}{c}{Feasibility} & \multicolumn{3}{c}{Computational efficiency} \\
\cmidrule(lr){2-3}\cmidrule(lr){4-6} 
& Global & \COMPASSicon COMPASS & Oracle Search (s) & \COMPASSicon COMPASS (s) & Speedup \\
\midrule
Simulated Annealing & 2,059 (68.63\%) & 2,994 (99.80\%) & 9.150 & 0.186 & 49.14\(\times\) \\
Bayesian Optimization & 2,669 (88.97\%) & 2,995 (99.83\%) & 168.221 & 0.291 & 577.72\(\times\) \\
\bottomrule
\end{tabular*}
\end{table}

\begin{table}[h!t]
\centering
\caption{Policy recovery performance of \COMPASSicon COMPASS by dimension under alternative search methods.}
\label{tab:search_compass_policy_recovery}
\begin{tabular*}{\textwidth}{
@{\extracolsep{\fill}}lcc@{}}
\toprule
Policy dimension & Simulated Annealing & Bayesian Optimization \\
\midrule
Demographic equity & 2,212 (73.73\%) & 2,776 (92.53\%) \\
Capacity shortage & 2,583 (86.10\%) & 2,762 (92.07\%) \\
Travel distance & 2,460 (82.00\%) & 2,558 (85.27\%) \\
School quality & 2,269 (75.63\%) & 2,573 (85.77\%) \\
Student balance & 1,397 (46.57\%) & 1,812 (60.40\%) \\
\textit{Composite policy score} & \textit{1,845 (61.50\%)} & \textit{2,443 (81.43\%)} \\
\bottomrule
\end{tabular*}
\end{table}

Changing the Oracle Search teacher substantially affects parameter-level prediction, particularly for the raw Difficulty-aware predictions. However, this variation did not translate into a loss of deployment feasibility. COMPASS achieved solver-feasible success rates of 99.80\% with Simulated Annealing supervision and 99.83\% with Bayesian Optimization supervision. Both settings also retained meaningful policy recovery across all five dimensions and the composite policy score and substantial speedups over their corresponding search procedures. Overall, the results show that COMPASS's deployment-level effectiveness is not restricted to supervision generated by a particular Oracle Search method.

\section{Sensitivity to Learning Models}
\label{sec:sensitivity_to_learning_models}

Having examined robustness to the source of Oracle Search supervision, we next considered whether the results depend on the particular model used to learn from that supervision. The primary implementation of COMPASS uses XGBoost within its Parameter Generator, which leaves open the possibility that the reported performance is specific to its inductive bias. We therefore instantiated the Parameter Generator using XGBoost, LightGBM, and Random Forest, while retaining the same optimization-ready urban representation, downstream zoning MILP, policy specification, and evaluation criteria. The three learning models are compared in Tables \ref{tab:learning_parameter_prediciton}, \ref{tab:learning_feasibility_runtime}, and \ref{tab:learning_compass_policy_recovery}, in terms of parameter prediction, feasibility progression, computational efficiency, and fixed-denominator policy recovery.

\begin{table}[h!t]
\centering
\caption{Parameter prediction performance across alternative learning models.}
\label{tab:learning_parameter_prediciton}
\begin{tabular*}{\textwidth}{
@{\extracolsep{\fill}}lcccccc@{}}
\toprule
\multirow{2}{*}{Learning model} & \multicolumn{3}{c}{Global} & \multicolumn{3}{c}{Difficulty-aware} \\
\cmidrule(lr){2-4}\cmidrule(lr){5-7} 
& MAE & RMSE & \(R^2\) & MAE & RMSE & \(R^2\) \\
\midrule
XGBoost & 0.529413 & 0.834521 & 0.578509 & 0.538910 & 0.985823 & 0.360189 \\
LightGBM & 0.561394 & 0.918661 & 0.508437 & 0.581094 & 1.074086 & 0.272197 \\
Random Forest & 0.552629 & 0.879556 & 0.537118 & 0.534567 & 0.929511 & 0.474510 \\
\bottomrule
\end{tabular*}
\end{table}

\begin{table}[h!t]
\centering
\caption{Feasibility and runtime performance across alternative learning models.}
\label{tab:learning_feasibility_runtime}
\begin{tabular*}{\textwidth}{
@{\extracolsep{\fill}}lccccc@{}}
\toprule
\multirow{2}{*}{Learning model} & \multicolumn{2}{c}{Feasibility} & \multicolumn{3}{c}{Computational efficiency} \\
\cmidrule(lr){2-3}\cmidrule(lr){4-6} 
& Global & \COMPASSicon COMPASS & Oracle Search (s) & \COMPASSicon COMPASS (s) & Speedup \\
\midrule
XGBoost & 2,059 (68.63\%) & 2,994 (99.80\%) & 9.150 & 0.186 & 49.14\(\times\) \\
LightGBM & 2,016 (67.20\%) & 2,992 (99.73\%) & 9.150 & 0.372 & 24.57\(\times\) \\
Random Forest & 2,101 (70.03\%) & 2,991 (99.70\%) & 9.150 & 0.392 & 23.34\(\times\) \\
\bottomrule
\end{tabular*}
\end{table}

\begin{table}[h!t]
\centering
\caption{Policy recovery performance of \COMPASSicon COMPASS by dimension across alternative learning models.}
\label{tab:learning_compass_policy_recovery}
\begin{tabular*}{\textwidth}{
@{\extracolsep{\fill}}lccc@{}}
\toprule
Policy dimension & XGBoost & LightGBM & Random Forest \\
\midrule
Demographic equity & 2,212 (73.73\%) & 2,202 (73.40\%) & 2,205 (73.50\%) \\
Capacity shortage & 2,583 (86.10\%) & 2,581 (86.03\%) & 2,586 (86.20\%) \\
Travel distance & 2,460 (82.00\%) & 2,445 (81.50\%) & 2,459 (81.97\%) \\
School quality & 2,269 (75.63\%) & 2,289 (76.30\%) & 2,267 (75.57\%) \\
Student balance & 1,397 (46.57\%) & 1,387 (46.23\%) & 1,398 (46.60\%) \\
\textit{Composite policy score} & \textit{1,845 (61.50\%)} & \textit{1,832 (61.07\%)} & \textit{1,849 (61.63\%)}\\
\bottomrule
\end{tabular*}
\end{table}

Although parameter-prediction accuracy varied across the three learning models, their downstream behavior was highly consistent. The Global ablation achieved solver-feasible success rates of 67.20\% - 70.03\%, which increased to 92.27\% - 92.93\% under Difficulty-aware Parameter Generation and further to 99.70\% - 99.80\% for COMPASS. Fixed-denominator recovery across the five policy dimensions and the composite policy score was also nearly identical among the three learning models. In addition, all variants retained substantial computational advantages over Oracle Search, with speedups ranging from 23.34\(\times\) to 49.14\(\times\). These findings demonstrate that COMPASS achieves consistent performance across the three evaluated tree-based learning models, indicating that its effectiveness is not specific to XGBoost.

\section{Sensitivity to Policy Preferences}
\label{sec:sensitivity_to_policy_preferences}

Finally, we moved from methodological choices to the policy context in which COMPASS is deployed. In the main experiments, search-derived supervision is generated and COMPASS is trained and evaluated under a diversity-prioritized policy specification, whose objective places strong emphasis on balancing FRL, focal-group, and diversity-index deviations across zones. Because the main evidence is obtained under this single equity-oriented specification, it remains unclear whether the framework is equally effective when policy priorities are changed. We therefore re-instantiated the complete framework under two alternative settings: a compactness- and quality-prioritized specification and a more strongly diversity- and capacity-prioritized specification. The detailed objective coefficients for the three policy specifications are reported in Table \ref{tab:policy_preference_parameter}. These settings modify the relative policy priorities while retaining the same COMPASS framework and evaluation criteria. Parameter prediction, solver-feasible MILP performance and computational efficiency, and fixed-denominator policy recovery across the three specifications are compared respectively in Tables \ref{tab:policy_parameter_prediciton}, \ref{tab:policy_feasibility_runtime}, and \ref{tab:policy_compass_policy_recovery}.

\begin{table}[h!t]
\centering
\caption{Policy-objective configurations for corresponding policy-preference specifications.}
\label{tab:policy_preference_parameter}
\begin{tabular*}{\textwidth}{
@{\extracolsep{\fill}}lccccc@{}}
\toprule
Policy preference & \(w_d\) & \(p_d\) & \(w_{div}\) & \(w_q\) & \(w_s\) \\
\midrule
Diversity & 0.10 & 2.0 & 75.0 & 1.0 & 2.0 \\
Compactness \& Quality & 0.20 & 4.0 & 60.0 & 1.5 &2.0 \\
Stronger diversity \& Capacity & 0.05 & 1.0 & 100.0 & 0.5 & 2.5 \\
\bottomrule
\end{tabular*}
\vspace{2pt}
\raggedright
\footnotesize
\(w_d\), \(w_{div}\), \(w_q\), \(w_s\): coefficients for distance, diversity, school quality, minimum surplus deviation; 
\(p_d\): exponent of distance term.
\end{table}

\begin{table}[h!t]
\centering
\caption{Parameter-prediction performance under alternative policy-preference specifications.}
\label{tab:policy_parameter_prediciton}
\begin{tabular*}{\textwidth}{
@{\extracolsep{\fill}}lcccccc@{}}
\toprule
\multirow{2}{*}{Policy preference} & \multicolumn{3}{c}{Global} & \multicolumn{3}{c}{Difficulty-aware} \\
\cmidrule(lr){2-4}\cmidrule(lr){5-7} 
& MAE & RMSE & \(R^2\) & MAE & RMSE & \(R^2\) \\
\midrule
Diversity & 0.529413 & 0.834521 & 0.578509 & 0.538910 & 0.985823 & 0.360189 \\
Compactness \& Quality & 0.654369 & 0.921258 & 0.525491 & 0.809559 & 1.351242 & 0.053373 \\
Stronger diversity \& Capacity & 0.710484 & 1.600649 & 0.360376 & 0.862932 & 1.992773 & 0.164570 \\
\bottomrule
\end{tabular*}
\end{table}

\begin{table}[h!t]
\centering
\caption{Solver-feasible MILP and runtime performance under alternative policy-preference specifications.}
\label{tab:policy_feasibility_runtime}
\begin{tabular*}{\textwidth}{
@{\extracolsep{\fill}}lccccc@{}}
\toprule
\multirow{2}{*}{Policy preference} & \multicolumn{2}{c}{Feasibility} & \multicolumn{3}{c}{Computational efficiency} \\
\cmidrule(lr){2-3}\cmidrule(lr){4-6} 
& Global & \COMPASSicon COMPASS & Oracle Search (s) & \COMPASSicon COMPASS (s) & Speedup \\
\midrule
Diversity & 68.63\% & 99.80\% & 9.150 & 0.186 & 49.14\(\times\) \\
Compactness \& Quality & 86.67\% & 99.93\% & 7.807 & 0.274 & 28.45\(\times\) \\
Stronger diversity \& Capacity & 86.67\% & 99.67\% & 7.141 & 0.336 & 21.26\(\times\) \\
\bottomrule
\end{tabular*}
\end{table}

\begin{table}[h!t]
\centering
\caption{Policy recovery of \COMPASSicon COMPASS by dimension under alternative policy-preference specifications.}
\label{tab:policy_compass_policy_recovery}
\begin{tabular*}{\textwidth}{
@{\extracolsep{\fill}}lccc@{}}
\toprule
Policy dimension & Diversity & Compactness \& Quality & Stronger diversity \& Capacity \\
\midrule
Demographic equity & 2,212 (73.73\%) & 2,554 (85.13\%) & 2,725 (90.83\%) \\
Capacity shortage & 2,583 (86.10\%) & 2,712 (90.40\%) & 2,756 (91.87\%) \\
Travel distance & 2,460 (82.00\%) & 2,529 (84.30\%) & 2,521 (84.03\%) \\
School quality & 2,269 (75.63\%) & 2,458 (81.93\%) & 2,528 (84.27\%) \\
Student balance & 1,397 (46.57\%) & 1,729 (57.63\%) & 1,823 (60.77\%) \\
\textit{Composite policy score} & \textit{1,845 (61.50\%)} & \textit{2,193 (73.10\%)} & \textit{2,245 (74.83\%)}\\
\bottomrule
\end{tabular*}
\end{table}

The policy specifications produced noticeable differences in parameter-level prediction accuracy, indicating that some policy configurations are more difficult to learn than others. Nevertheless, COMPASS achieved solver-feasible success rates between 99.67\% and 99.93\% across all three settings. Fixed-denominator recovery also remained substantial across the five policy dimensions and the composite policy score, although student balance consistently exhibited the lowest recovery rate. Moreover, COMPASS remained 21.26\(\times\) - 49.14\(\times\) faster than the corresponding Oracle Search. These results demonstrate that the effectiveness of COMPASS is not confined to the default policy specification.

\section{Additional Case Studies}
\label{sec:additional_case_studies}

Aggregate feasibility and policy-gap metrics characterize the overall performance of COMPASS, but they do not show how the resulting zoning plans are expressed spatially. Therefore, ten illustrative out-of-fold city cases are selected to span distinct feasibility pathways in Figure \ref{fig:ZoneOptimization_Figure_3}: three shared-feasibility cases, three Difficulty-aware recovery cases, two COMPASS-only recovery cases, and two unresolved hard cases. For each city, the columns show the processed city context before zoning, the Oracle Search reference, and the zoning plans generated by the Global ablation, the Difficulty-aware ablation, and COMPASS, respectively. In the three learning-based columns, an empty cell indicates that the corresponding pipeline did not produce a solver-verified feasible zoning plan within the evaluation budget.

\begin{figure}[p]
\centering
\includegraphics[width=0.91\textwidth,keepaspectratio]{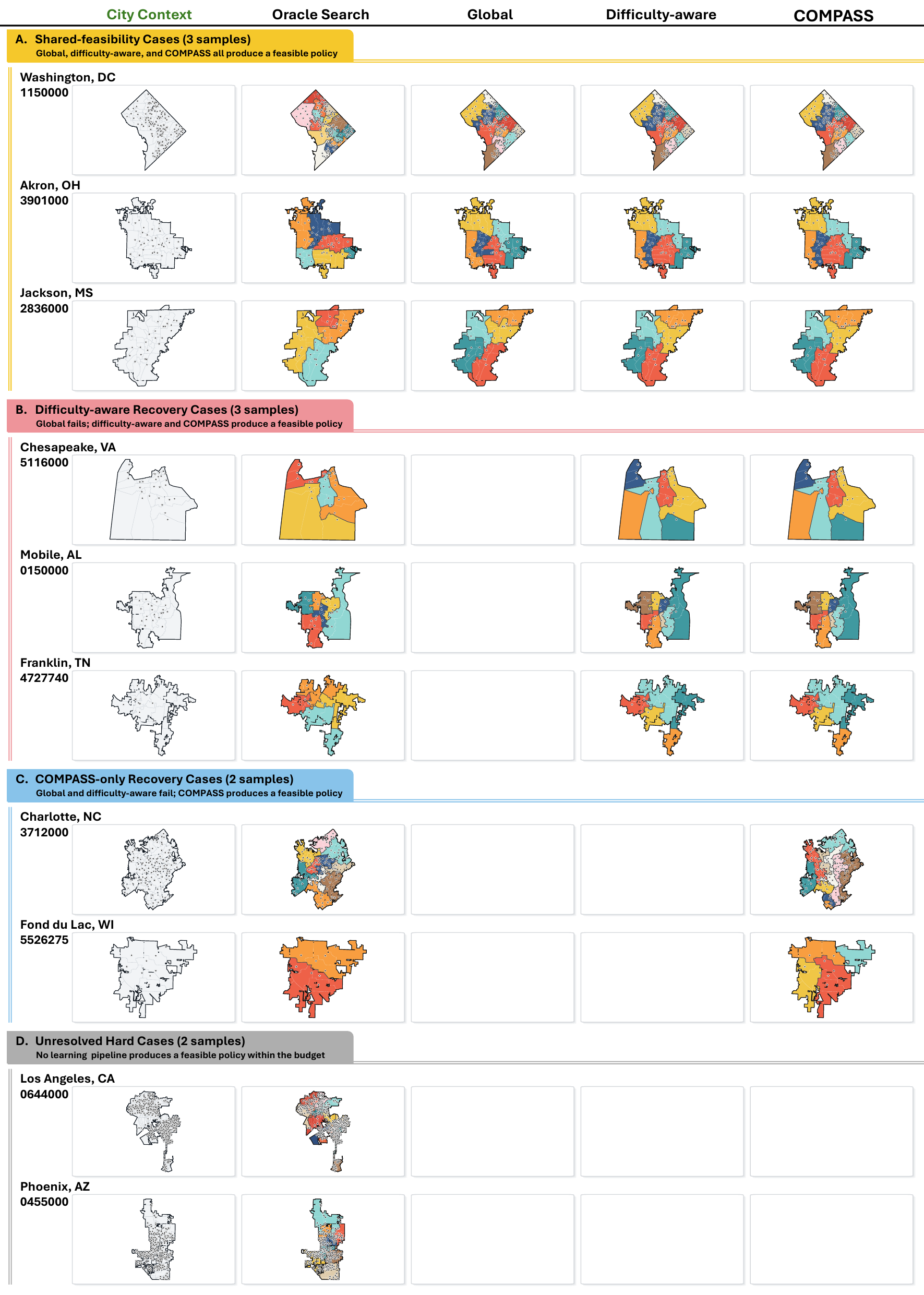}
\caption{(\textit{A}) Shared feasibility in Washington, D.C.; Akron, Ohio; and Jackson, Mississippi. (\textit{B}) Difficulty-aware recovery after failure of the Global ablation in Chesapeake, Virginia; Mobile, Alabama; and Franklin, Tennessee. (\textit{C}) COMPASS-only recovery through Conditional Rescue in Charlotte, North Carolina, and Fond du Lac, Wisconsin. (\textit{D}) Unresolved cases in Los Angeles, California, and Phoenix, Arizona, with Oracle Search references available.}
\label{fig:ZoneOptimization_Figure_3}
\end{figure}

In the shared-feasibility cases, COMPASS is not expected to reproduce the search-derived boundaries exactly; what matters is whether it preserves the resulting policy outcomes. The Difficulty-aware recovery cases show that incorporating city-specific optimization difficulty can recover feasible zoning plans when the Global ablation fails. As demonstrated by the COMPASS-only cases, Conditional Rescue can further extend feasible coverage to harder cities, although this extension may come at the cost of larger policy gaps. The unresolved cases delineate the current boundary of the framework: a small number of large cities with complex transportation structures may remain unsolved within the evaluation budget because of time limits or persistent infeasibility. These cases provide only spatial case-level interpretations, whereas the statistical conclusions on feasibility and policy recovery are based on the full 3,000-city evaluation rather than these selected examples.

\section{Dataset and Additional Methodological Specifications}
\label{sec:dataset_and_additional_methodological_specifications}

This section provides the dataset and methodological specifications omitted from the main text for conciseness. We first describe the construction and provenance of the nationwide city dataset, followed by the complete feature dictionary for the 140-dimensional optimization-ready urban representation. We then specify the construction of the cross-city supervision archive and the symmetric Global reference objective. Finally, we report the exact definitions of the five structural difficulty dimensions and the search-supervised plan-generation difficulty label, the difficulty-sensitive asymmetric learning weights, the deterministic parameter adjustments applied before the initial MILP solve, and the Conditional Rescue rules. The materials presented here consist of data definitions, fixed constants, mathematical specifications, and decision rules rather than additional experimental results.

\paragraph{Dataset overview.} The unit of analysis is a US Census PLACE-level city instance \(C_i\), and the dataset integrates several national data sources, as is summarized in Table \ref{tab:national_data_source}. For readability, we refer to each instance as a city, although Census PLACE geography includes both incorporated places and census-designated places. The retained dataset contains 3,000 unique PLACE instances spanning all 50 states and the District of Columbia. An instance was retained only if it had a valid PLACE geometry, nonempty intersecting Census blocks with demographic information, nonempty level-1 elementary attendance-zone coverage, and at least 2 eligible elementary-serving schools both within the effective city geometry and after spatial matching to the processed attendance units. Eligible schools were regular public schools that were open or newly opened, were not fully virtual, served at least one grade from pre-kindergarten through Grade 5, and had valid geographic coordinates. Every retained instance consequently contained at least two usable optimization units after spatial processing and school matching.

\begin{table}[h!t]
\centering
\caption{Data sources used to construct the nationwide school-zoning dataset.}
\label{tab:national_data_source}
\begin{tabular*}{\textwidth}{
@{\extracolsep{\fill}}ll@{}}
\toprule
Data component & Source \\
\midrule
City boundaries & US Census TIGER/Line PLACE Files \\
Spatial and demographic data & US Census Tabulation Blocks and PL 94-171 Data \\
Elementary attendance zones & NCES School Attendance Boundary Survey Primary-Level Files \\
School characteristics & NCES Common Core of Data \\
School quality & EDFacts FS175 and FS178 Files \\
\bottomrule
\end{tabular*}
\end{table}

We constructed spatial processing rules that materially affected the geographic extent or optimization units of an instance only. For a multipart PLACE geometry, a component was considered supported when it contained positive block population and at least one eligible school with positive enrollment. When exactly one supported component accounted for at least 90\% of the total PLACE area and was at least ten times larger than the largest unsupported component, only the dominant component was retained. Otherwise, school-less components were removed when they accounted for less than 1\% of the total area or were located at least 100 \(\mathrm{m}\) from every supported component. Instances that subsequently retained multiple disconnected supported components or unsupported empty components were excluded. Elementary attendance zones were then clipped to the effective PLACE geometry. When the ratio between the summed clipped-zone area and their union area exceeded 1.10, the zone and city boundaries were polygonized into mutually exclusive cells, with cells no larger than 1 \(\mathrm{m}^2\) removed. Schools and Census blocks were assigned to the resulting units spatially, with nearest-unit assignment used only for boundary cases.

\paragraph{140-dimensional optimization-ready urban representation.} The optimization-ready encoder maps each heterogeneous city instance to a fixed-dimensional vector, \(\mathbf{x}_i=\mathcal{E}(C_i)\in\mathbb{R}^{d}\), with \(d=140\) in the current implementation, thereby providing a common numerical interface for a single Parameter Generator across cities. The representation consists of five optimization-relevant blocks: 18 spatial-structure features, 39 school-supply features, 10 student-demand features, 27 demographic-equity features, and 46 optimization-context features. The complete feature dictionary is reported in Tables \ref{tab:spatial_structure_18dimension}, \ref{tab:school_supply_39dimension}, \ref{tab:student_demand_10dimension}, \ref{tab:demographic_equity_27dimension}, and \ref{tab:optimization_context_46dimension}.

\begin{table}[h!t]
\centering
\caption{Spatial structure \(\mathbf{x}_i^{\mathrm{spatial}}\) - 18 dimensions.}
\label{tab:spatial_structure_18dimension}
\small
\begin{tabularx}{\textwidth}{@{}lc>{\raggedright\arraybackslash}X@{}}
\toprule
Feature & Dimension & Definition \\
\midrule
unit\_count & 1 & Number of mutually exclusive geographic units after spatial processing \\
graph\_node\_count & 1 & Number of nodes in the processed spatial-adjacency graph \\
graph\_edge\_count & 1 & Number of undirected adjacency edges \\
graph\_component\_count & 1 & Number of connected components in the adjacency graph \\
graph\_density & 1 & Ratio of observed adjacency edges to all possible undirected edges \\
graph\_isolate\_count & 1 & Number of isolated assignment units \\
zone\_2suffix & 2 & City-specific lower and upper bounds on the admissible number of zones \\
city\_distance\_scale\_miles & 1 & Characteristic citywide distance scale used to realize the spatial-accessibility factor \\
milp\_unit\_count & 1 & Number of geographic assignment units supplied to the zoning MILP \\
milp\_neighbor\_link\_count & 1 & Number of neighboring-unit links supplied to the zoning MILP \\
graph\_degree\_7suffix & 7 & Distributional summaries of node degree in the spatial-adjacency graph \\
\bottomrule
\end{tabularx}
\vspace{2pt}
\raggedright
\footnotesize
2suffix:\{min, max\};
7suffix: \{sum, mean, std, min, max, q25, q75\}.
\end{table}

\begin{table}[h!t]
\centering
\caption{School supply \(\mathbf{x}_i^{\mathrm{school}}\) - 39 dimensions.}
\label{tab:school_supply_39dimension}
\small
\begin{tabularx}{\textwidth}{@{}lc>{\raggedright\arraybackslash}X@{}}
\toprule
Feature & Dimension & Definition \\
\midrule
total\_schools & 1 & Number of eligible and spatially matched elementary public schools \\
school\_unit\_count & 1 & Number of assignment units containing at least one eligible school \\
capacity\_student\_ratio & 1 & Ratio of the enrollment-based capacity proxy to student demand \\
school\_count\_9suffix & 9 & Distributional summaries of eligible-school counts across assignment units \\
block\_ele\_cap\_9suffix & 9 & Distributional summaries of the capacity proxy across assignment units \\
school\_quality\_index\_9suffix & 9 & Distributional summaries of the unit-level school-quality index \\
school\_quality\_weight\_9suffix & 9 & Distributional summaries of the support weight for school quality \\
\bottomrule
\end{tabularx}
\vspace{2pt}
\raggedright
\footnotesize
9suffix: \{sum, mean, std, min, max, q25, q75, weighted\_mean, weighted\_std\}.
\end{table}

\begin{table}[h!t]
\centering
\caption{Student demand \(\mathbf{x}_i^{\mathrm{student}}\) - 10 dimensions.}
\label{tab:student_demand_10dimension}
\small
\begin{tabularx}{\textwidth}{@{}lc>{\raggedright\arraybackslash}X@{}}
\toprule
Feature & Dimension & Definition \\
\midrule
total\_students & 1 & Total student demand represented using the child-population proxy \\
students\_9suffix & 9 & Distributional summaries of student demand across assignment units \\
\bottomrule
\end{tabularx}
\vspace{2pt}
\raggedright
\footnotesize
9suffix: \{sum, mean, std, min, max, q25, q75, weighted\_mean, weighted\_std\}.
\end{table}

\begin{table}[h!t]
\centering
\caption{Demographic equity \(\mathbf{x}_i^{\mathrm{equity}}\) - 27 dimensions.}
\label{tab:demographic_equity_27dimension}
\small
\begin{tabularx}{\textwidth}{@{}lc>{\raggedright\arraybackslash}X@{}}
\toprule
Feature & Dimension & Definition \\
\midrule
FRL\_pct\_9suffix & 9 & Distributional summaries of the unit-level free or reduced-price lunch share \\
focal\_group\_pct\_9suffix & 9 & Distributional summaries of the combined share of multiracial populations \\
diversity\_index\_9suffix & 9 & Distributional summaries of the unit-level diversity index \\
\bottomrule
\end{tabularx}
\vspace{2pt}
\raggedright
\footnotesize
9suffix: \{sum, mean, std, min, max, q25, q75, weighted\_mean, weighted\_std\}.
\end{table}

\begin{table}[h!t]
\centering
\caption{Optimization context \(\mathbf{x}_i^{\mathrm{context}}\) - 46 dimensions.}
\label{tab:optimization_context_46dimension}
\small
\begin{tabularx}{\textwidth}{@{}lc>{\raggedright\arraybackslash}X@{}}
\toprule
Feature & Dimension & Definition \\
\midrule
annealing\_iterations & 1 & Number of iterations allocated to Oracle Search \\
annealing\_initial\_temperature & 1 & Initial temperature of the simulated-annealing search \\
annealing\_cooling\_rate & 1 & Cooling rate of the simulated-annealing search \\
annealing\_seed & 1 & Random seed used by Oracle Search \\
annealing\_2suffix\_zones & 2 & Global lower and upper bounds on the zone-count domain \\
annealing\_balance\_tolerance\_2suffix & 2 & Search-domain bounds for enrollment-balance tolerance \\
annealing\_cover\_distance\_factor\_2suffix & 2 & Search-domain bounds for the spatial-accessibility factor \\
annealing\_max\_shortage\_fraction\_2suffix & 2 & Search-domain bounds for the capacity-shortage fraction \\
annealing\_school\_slack\_below\_2suffix & 2 & Search-domain bounds for the downward school-distribution slack \\
annealing\_school\_slack\_above\_2suffix & 2 & Search-domain bounds for the upward school-distribution slack \\
annealing\_frl\_lower\_2suffix & 2 & Search-domain bounds for the lower FRL bound \\
annealing\_frl\_upper\_2suffix & 2 & Search-domain bounds for the upper FRL bound \\
annealing\_focal\_group\_lower\_2suffix & 2 & Search-domain bounds for the lower focal-group bound \\
annealing\_focal\_group\_upper\_2suffix & 2 & Search-domain bounds for the upper focal-group bound \\
annealing\_diversity\_index\_lower\_2suffix & 2 & Search-domain bounds for the lower diversity-index bound \\
annealing\_diversity\_index\_upper\_2suffix & 2 & Search-domain bounds for the upper diversity-index bound \\
annealing\_school\_quality\_lower\_2suffix & 2 & Search-domain bounds for the lower school-quality bound \\
annealing\_school\_quality\_upper\_2suffix & 2 & Search-domain bounds for the upper school-quality bound \\
policy\_objective\_quality\_weight & 1 & Objective weight assigned to school quality \\
policy\_objective\_distance\_weight & 1 & Objective weight assigned to travel distance \\
policy\_objective\_distance\_power & 1 & Objective exponent applied to the distance component \\
policy\_objective\_diversity\_weight & 1 & Objective weight assigned to demographic diversity \\
policy\_objective\_min\_surplus\_weight & 1 & Objective weight assigned to minimum capacity surplus \\
score\_weight\_student\_cv & 1 & Evaluation weight for variation in student enrollment \\
score\_weight\_school\_dev & 1 & Evaluation weight for deviations in school distribution \\
score\_weight\_frl\_spread & 1 & Evaluation weight for between-zone FRL disparity \\
score\_weight\_focal\_group\_spread & 1 & Evaluation weight for between-zone focal-group disparity \\
score\_weight\_diversity\_spread & 1 & Evaluation weight for between-zone diversity disparity \\
score\_weight\_quality\_spread & 1 & Evaluation weight for between-zone school-quality disparity \\
score\_weight\_shortage & 1 & Evaluation weight for capacity shortage \\
score\_weight\_distance & 1 & Evaluation weight for travel distance \\
score\_weight\_disconnected & 1 & Evaluation penalty for disconnected zones \\
\bottomrule
\end{tabularx}
\vspace{2pt}
\raggedright
\footnotesize
2suffix: \{min, max\}.
\end{table}

\paragraph{Oracle Search supervision and the symmetric Global ablation.}
For each training city \(i\), Oracle Search evaluates a finite set of candidate control vectors \(\mathcal{T}_i^{S}\subseteq\Theta\). The Oracle Search teacher control vector is defined as the feasible candidate with the lowest recorded policy objective among the evaluated candidates:
\begin{equation}
\label{eq:search_teacher_selection}
\boldsymbol{\theta}_i^{S}\in\arg\min_{\boldsymbol{\theta}\in\mathcal{T}_i^{S}:\boldsymbol{\pi}_i^{\mathrm{MILP}}(\boldsymbol{\theta},\boldsymbol{\rho})\neq\varnothing}J_i\!\left(\boldsymbol{\pi}_i^{\mathrm{MILP}}(\boldsymbol{\theta},\boldsymbol{\rho});\boldsymbol{\rho}\right).
\end{equation}
Here, \(\boldsymbol{\pi}_i^{\mathrm{MILP}}(\boldsymbol{\theta},\boldsymbol{\rho})\) denotes the zoning plan returned by the downstream MILP under control vector \(\boldsymbol{\theta}\) and fixed policy specification \(\boldsymbol{\rho}\). Accordingly, \(\boldsymbol{\theta}_i^{S}\) is the best feasible candidate encountered within the finite evaluated set and is not claimed to be a global optimum over \(\Theta\). The resulting cross-city supervision archive is \(\mathcal{D}^{S}=\left\{\left(\mathbf{x}_i,\boldsymbol{\theta}_i^{S}\right)\right\}_{i=1}^{N}\), where \(\mathbf{x}_i\in\mathbb{R}^{140}\) is the optimization-ready urban representation of city \(i\).

The Global ablation provides the symmetric reference model used before introducing difficulty-sensitive weighting. It predicts \(\widehat{\boldsymbol{\theta}}_i^{G}=g_{\widehat{\boldsymbol{\phi}}_G}(\mathbf{x}_i)\) and minimizes the equally weighted parameter-level squared error:
\begin{equation}
\label{eq:global_symmetric_objective}
\widehat{\boldsymbol{\phi}}_G=\arg\min_{\boldsymbol{\phi}_G}\frac{1}{Np}\sum_{i=1}^{N}\sum_{j=1}^{p}\left(\widehat{\theta}_{ij}^{G}-\theta_{ij}^{S}\right)^2,\quad p=14.
\end{equation}
The difficulty-sensitive asymmetric learning objective defined below retains the same supervision target and output interface but changes how prediction errors are weighted across cities, parameters, and error directions.

\paragraph{Difficulty-signal construction.} The complete difficulty signals \(\boldsymbol{\delta}_i\) consist of a structural difficulty profile \(\boldsymbol{\delta}_i^{\mathrm{str}}\) and a plan-generation difficulty label \(\delta_i^{\mathrm{gen}}\). Before constructing these signals, all contributing variables are placed on a comparable scale. For city \(i\) and feature \(m\), we robustly standardize \(x_{im}\) as
\begin{equation}
\label{eq:robust_standardization}
\operatorname{rz}_{\mathcal I_{\mathrm{ref}}}(x_{im})=\operatorname{clip}\left(\frac{x_{im}-\operatorname{median}_{r\in\mathcal I_{\mathrm{ref}}}(x_{rm})}{\operatorname{scale}_{\mathcal I_{\mathrm{ref}}}(x_m)},-6,6\right).
\end{equation}
Here, \(r\) indexes a city in the reference set. The set \(\mathcal I_{\mathrm{ref}}\subseteq\{1,\ldots,N\}\) contains the city indices used to estimate the normalization statistics and the subsequent discretization thresholds. The quantity \(\operatorname{median}_{r\in\mathcal I_{\mathrm{ref}}}(x_{rm})\) is the reference-set median of feature \(m\). Any missing value of \(x_{im}\) is replaced by this median before robust standardization is applied. The feature-specific scaling term in the denominator is defined as
\begin{equation}
\label{eq:feature_specific_scale}
\operatorname{scale}_{\mathcal I_{\mathrm{ref}}}(x_m)=\begin{cases}
\operatorname{IQR}_{\mathcal I_{\mathrm{ref}}}(x_m),&\operatorname{IQR}_{\mathcal I_{\mathrm{ref}}}(x_m)>10^{-9},\\ 
\operatorname{sd}_{0,\mathcal I_{\mathrm{ref}}}(x_m),&\operatorname{IQR}_{\mathcal I_{\mathrm{ref}}}(x_m)\leq10^{-9}\ \text{and}\ \operatorname{sd}_{0,\mathcal I_{\mathrm{ref}}}(x_m)>10^{-9},\\ 
1,&\text{otherwise},
\end{cases}
\end{equation}
where \(\operatorname{IQR}_{\mathcal I_{\mathrm{ref}}}(x_m)\) is the difference between the 75th and 25th percentiles of feature \(m\) in the reference set, and \(\operatorname{sd}_{0,\mathcal I_{\mathrm{ref}}}(x_m)\) is its population standard deviation. The threshold \(10^{-9}\) identifies numerically degenerate dispersion. Thus, the interquartile range is used whenever it is nondegenerate, the population standard deviation is used when the interquartile range is degenerate but the standard deviation is not, and a unit scale is used when both are degenerate. Finally, we use \(\operatorname{clip}(\cdot,-6,6)\) to restrict the standardized value to the interval \([-6,6]\).

On this common robust standardization, five structural scores are constructed from the optimization-ready urban representation:
\begin{equation}
\label{eq:structure_score}
\xi_i^k=\sum_{m\in\mathcal F_k}a_{km}\operatorname{rz}(x_{im}),\quad k\in\left\{\mathrm{spatial},\mathrm{capacity},\mathrm{school},\mathrm{demographic},\mathrm{graph}\right\},
\end{equation}
where \(\mathcal F_k\) denotes the feature subset associated with structural dimension \(k\), and \(a_{km}\in\{-1,+1\}\) specifies the direction of the contribution of feature \(m\). The spatial-service score increases with the citywide distance scale, the processed spatial-unit count, and the number of units entering the MILP, but decreases with graph density and mean node degree. The capacity-tightness score increases with the maximum and student-weighted mean of both student demand and the enrollment-based capacity proxy, while decreasing with the aggregate capacity-to-demand ratio. The school-distribution-imbalance score combines the standard deviation, maximum, and student-weighted standard deviation of unit-level student demand and school counts, with all terms entering positively. The demographic-constraint-risk score similarly combines dispersion in FRL share, focal-group share, racial diversity, and school quality. The graph-fragmentation-risk score increases with the numbers of connected components and isolated units and with degree variation, but decreases with graph density and mean degree. These scores are discretized at their respective 75th percentiles over the reference-city set. Let \(b_{0.75}^{k}\) denote the corresponding cut point for dimension \(k\). For the 3,000-city reference corpus used in the reported experiments, the resulting cut points were
\begin{equation}
\label{eq:experiment_cut_point}
\bigl(b_{0.75}^{\mathrm{spatial}},~b_{0.75}^{\mathrm{capacity}},~b_{0.75}^{\mathrm{school}},~b_{0.75}^{\mathrm{demographic}},~b_{0.75}^{\mathrm{graph}}\bigr)=\bigl(1.897112,\,2.571794,\,3.134181,\,1.990016,\,0.926073\bigr).
\end{equation}
All five structural labels are then assigned according to
\begin{equation}
\label{eq:structural_classification}
\begin{aligned}
\delta_i^k&=\begin{cases}\mathrm{High}, & \xi_i^k\geq b_{0.75}^k,\\\mathrm{Low},  & \text{otherwise},\end{cases}\quad k\in\{\mathrm{spatial},\mathrm{school},\mathrm{demographic}\},\\ 
\delta_i^{\mathrm{capacity}}&=\begin{cases}\mathrm{Tight},&\xi_i^{\mathrm{capacity}}\geq b_{0.75}^{\mathrm{capacity}},\\\mathrm{Loose},&\text{otherwise},\end{cases}\\ 
\delta_i^{\mathrm{graph}}&=\begin{cases}\mathrm{High},&\xi_i^{\mathrm{graph}}\geq b_{0.75}^{\mathrm{graph}}\ \lor\ x_{i,\texttt{graph\_component\_count}}>1\ \lor\ x_{i,\texttt{graph\_isolate\_count}}>0,\\\mathrm{Low},&\text{otherwise}.\end{cases}
\end{aligned}
\end{equation}

Together, these labels form the structural difficulty profile \(\boldsymbol{\delta}_i^{\mathrm{str}}=[\delta_i^{\mathrm{spatial}},\delta_i^{\mathrm{capacity}},\delta_i^{\mathrm{school}},\delta_i^{\mathrm{demographic}},\delta_i^{\mathrm{graph}}]\).

The search-supervised plan-generation difficulty label is obtained from a continuous teacher score constructed from the teacher control vector \(\boldsymbol{\theta}_i^S\) and the feasibility history recorded during Oracle Search. The continuous teacher score is defined as
\begin{equation}
\label{eq:continuous_teacher_score}
\begin{aligned}
\xi_i^{\mathrm{gen}}={}&\operatorname{rz}(\tau_i^S)+\operatorname{rz}(\kappa_i^S)+\operatorname{rz}(\eta_i^S)+\operatorname{rz}(s_i^{S,-})+\operatorname{rz}(s_i^{S,+})\\
&+\sum_{g\in\mathcal G}\left[\operatorname{rz}(u_{ig}^S)-\operatorname{rz}(\ell_{ig}^S)\right]+\operatorname{rz}(\mathrm{IC}_i)+\operatorname{rz}(\mathrm{IR}_i),
\end{aligned}
\end{equation}
where \(\mathrm{IC}_i\) denotes the number of exact-MILP candidate evaluations reported as infeasible up to and including the occurrence of the selected teacher state, and \(\mathrm{IR}_i\) denotes their proportion among the exact-MILP evaluations in the same prefix. These quantities characterize the feasibility history leading to the retained teacher rather than the complete diagnostics of all candidates evaluated during the full Oracle Search run. The positively signed terms capture the extent to which Oracle Search required wider enrollment-balance, spatial-accessibility, capacity-shortage, school-distribution, and upper-bound settings, together with the frequency of infeasible candidate evaluations. Lower-bound parameters enter negatively because decreasing them relaxes the corresponding constraints. The teacher zone count \(K_i^S\) is excluded because its effect on feasibility is not directionally monotone. To reduce the skewness of the teacher scores while preserving their signs and ordering, we apply the signed-log transformation
\begin{equation}
\label{eq:signed-log-transformation}
\widetilde{\xi}_i^{\mathrm{gen}}=\operatorname{sign}\!\left(\xi_i^{\mathrm{gen}}\right)\log\!\left(1+\left|\xi_i^{\mathrm{gen}}\right|\right).
\end{equation}
The transformed scores are subsequently partitioned using Jenks natural breaks. For the 3,000-city reference set used in the reported experiments, the three internal breakpoints were
\begin{equation}
\label{eq:internal_breakpoint}
(b_1,~b_2,~b_3)=(-0.927105,\,0.479153,\,2.592100).
\end{equation}
The search-supervised plan-generation difficulty label is then assigned according to
\begin{equation}
\label{eq:search_supervised_label}
\delta_i^{\mathrm{gen}}=\begin{cases}\mathrm{Easy},&\widetilde{\xi}_i^{\mathrm{gen}}\leq b_1,\\\mathrm{Moderate},&b_1<\widetilde{\xi}_i^{\mathrm{gen}}\leq b_2,\\\mathrm{Hard},&b_2<\widetilde{\xi}_i^{\mathrm{gen}}\leq b_3,\\\mathrm{Extreme},&\widetilde{\xi}_i^{\mathrm{gen}}>b_3.
\end{cases}
\end{equation}

For an unseen target city, the auxiliary classifier predicts this label as
\[
\widehat{\delta}_i^{\mathrm{gen}}=h_{\widehat{\boldsymbol{\psi}}}(\mathbf{x}_i)\in\{\mathrm{Easy},\mathrm{Moderate},\mathrm{Hard},\mathrm{Extreme}\}.
\]

\paragraph{Difficulty-sensitive asymmetric learning weights.} Let \(e_{ij}=\widehat{\theta}_{ij}^{D}-\theta_{ij}^{S}\) denote the raw prediction error for control parameter \(j\) in city \(i\), consistent with the asymmetric loss in the main text. The corresponding difficulty-sensitive weight is
\begin{equation}
\label{eq:asymmetric_weight}
w_{ij} = \lambda_{\text{safe}} + (\lambda_{\text{risk}} - \lambda_{\text{safe}})\mathbb{I}\{ q_je_{ij} > 0 \}, \quad\lambda_{\text{risk}} > \lambda_{\text{safe}} > 0
\end{equation}
and
\begin{equation}
\label{eq:asymmetric_objective}
\mathcal{L}_{\mathrm{asym}}=\frac{1}{Np}\sum_{i=1}^N\sum_{j=1}^p w_{ij}e_{ij}^2.
\end{equation}
The parameter-wise feasibility-risk directions and experimental weights used to instantiate this objective are specified below:
\begin{equation}
\label{eq:feasibility_risk_direction}
q_j=\begin{cases}0,&\theta_{ij}\equiv K_i,\\[1mm]-1,&\theta_{ij}\in\left\{\tau_i,\kappa_i,\eta_i,s_i^-,s_i^+,\{u_{ig}\}_{g\in\mathcal G}\right\},\\[1mm]+1,&\theta_{ij}\in\left\{\{\ell_{ig}\}_{g\in\mathcal G}\right\}.
\end{cases}
\end{equation}
The number of zones \(K_i\) was treated symmetrically. Underprediction was treated as feasibility-risky for the enrollment-balance tolerance \(\tau_i\), spatial-accessibility factor \(\kappa_i\), capacity-shortage allowance \(\eta_i\), school-distribution slacks \(s_i^-\) and \(s_i^+\), and all upper policy bounds \(\{u_{ig}\}_{g\in\mathcal G}\). Conversely, overprediction was treated as risky for the lower policy bounds \(\{\ell_{ig}\}_{g\in\mathcal G}\). With \(\lambda_{\mathrm{safe}}=1\) and \(\lambda_{\mathrm{risk}}=4\), a prediction error in the feasibility-risky direction receives four times the training weight of an error of the same magnitude in the relatively safe direction.

\paragraph{Difficulty-dependent parameter relaxation.} The raw control-vector prediction from Difficulty-aware Parameter Generation is converted into an initial MILP configuration through three deterministic transformations. First, the prediction is relaxed according to the predicted plan-generation difficulty label, \(\check{\boldsymbol{\theta}}_i=\mathcal P_{\widehat{\delta}_i^{\mathrm{gen}}}(\widehat{\boldsymbol{\theta}}_i^D)\). Second, validity projection enforces the common parameter domain \(\Theta\), \(\overline{\boldsymbol{\theta}}_i=\mathcal V_\Theta(\check{\boldsymbol{\theta}}_i)\). Finally, city-specific realization converts the admissible control vector into the absolute quantities and centers required by the zoning MILP, \(\boldsymbol{\omega}_i^{[0]}=\mathcal R(C_i,\overline{\boldsymbol{\theta}}_i)\).

Let \(c=\widehat{\delta}_i^{\mathrm{gen}}\) denote the predicted plan-generation difficulty label. The relaxation operator leaves the predicted zone count unchanged, while modifying the remaining parameters in feasibility-oriented directions. The parameter mappings are
\begin{equation}
\label{eq:parameter_mapping}
\begin{aligned}
\check K_i&=\widehat K_i^D,\\
\check{\tau}_i&=\operatorname{clip}_{[0,2.5]}\left(\alpha_c^\tau\widehat{\tau}_i^D+\beta_c^\tau\right),\\
\check{\kappa}_i&=\max\left(0,\alpha_c^\kappa\widehat{\kappa}_i^D\right),\\
\check{\eta}_i&=\operatorname{clip}_{[0,1.5]}\left(\alpha_c^\eta\widehat{\eta}_i^D+\beta_c^\eta\right),\\
\check{s}_i^\pm&=\max\left(0,\widehat{s}_i^{D,\pm}+\beta_c^s\right),\\
\check{\ell}_{ig}&=\max\left(0,\alpha_c^\ell\widehat{\ell}_{ig}^D-\beta_c^\ell\right),\\
\check{u}_{ig}&=\max\left(0,\alpha_c^u\widehat{u}_{ig}^D+\beta_c^u\right),\quad g\in\mathcal G. 
\end{aligned}
\end{equation}
Easy predictions are therefore unchanged before validity projection, whereas Moderate, Hard, and Extreme predictions receive progressively stronger relaxation. The complete coefficient schedule is reported in Table \ref{tab:parameter_relaxation}.

\begin{table}[h!t]
\centering
\caption{Multiplicative and additive coefficients for difficulty-dependent parameter relaxation by predicted plan-generation difficulty.}
\label{tab:parameter_relaxation}
\begin{tabular*}{\textwidth}{
@{\extracolsep{\fill}}lcccccccccc@{}}
\toprule
Difficulty \(c\) & \(\alpha^\tau\) & \(\beta^\tau\) & \(\alpha^\kappa\) & \(\alpha^\eta\) & \(\beta^\eta\) & \(\beta^s\) & \(\alpha^\ell\) & \(\beta^\ell\) & \(\alpha^u\) & \(\beta^u\) \\
\midrule
Easy & 1.00 & 0.00 & 1.00 & 1.00 & 0.00 & 0.00 & 1.00 & 0.00 & 1.00 & 0.00 \\
Moderate & 1.20 & 0.08 & 1.25 & 1.35 & 0.05 & 2.00 & 0.88 & 0.04 & 1.22 & 0.08 \\
Hard & 1.85 & 0.25 & 1.80 & 1.85 & 0.18 & 6.00 & 0.60 & 0.16 & 1.80 & 0.22 \\
Extreme & 2.20 & 0.40 & 2.25 & 2.20 & 0.25 & 10.00 & 0.45 & 0.25 & 2.20 & 0.35 \\
\bottomrule
\end{tabular*}
\end{table}

The relaxed vector may still contain nonintegral zone or slack values, negative quantities, values outside the admissible initial ranges, or reversed lower-upper pairs. The validity operator \(\mathcal V_\Theta\) corrects these parameter-level inconsistencies before city-specific realization. Let \(n_i^{\mathrm{school}}\) denote the number of eligible schools and recall that \(n_i\) is the number of assignment units. The projected parameter components are defined by
\begin{equation}
\label{eq:projected_parameter_component}
\begin{aligned}
\overline K_i&=\operatorname{round}[\operatorname{clip}\left(\check K_i, 1, \max\left(1, \min\left(n_i, n_i^{\mathrm{school}}\right)\right)\right)],\\
\overline{\tau}_i&=\operatorname{clip}\left(\check{\tau}_i, 0, 2.5\right),\\
\overline{\kappa}_i&=\max\left(0,\check{\kappa}_i\right),\\
\overline{\eta}_i&=\operatorname{clip}\left(\check{\eta}_i, 0, 1.5\right),\\
\overline{s}_i^\pm&=\operatorname{round}[\operatorname{clip}\left(\check{s}_i^\pm, 0, \max\left(20, n_i^{\mathrm{school}}\right)\right)],\\
\left(\overline{\ell}_{ig}, \overline{u}_{ig}\right)&=\operatorname{sort}\left(\max\left(0,\check{\ell}_{ig}\right), \max\left(0,\check{u}_{ig}\right)\right).\\
\end{aligned}
\end{equation}

The realization operator \(\mathcal R\) then combines the admissible control vector with the target city instance. Let \(n_i^{\mathrm{student}}\) denote total student demand and \(d_i^{\mathrm{city}}\) denote the city distance scale. The absolute enrollment-balance, spatial-accessibility, capacity-shortage, and school-distribution quantities supplied to the MILP are calculated as
\begin{equation}
\label{eq:absolute_supply_to_MILP}
\begin{aligned}
\operatorname{BalanceBudget}_i&=\overline{\tau}_i \frac{n_i^{\mathrm{student}}}{\overline K_i},\\
\operatorname{DistanceLimit}_i&=\overline{\kappa}_id_i^{\mathrm{city}},\\
\operatorname{ShortageBudget}_i&=\overline{\eta}_i\frac{n_i^{\mathrm{student}}}{\overline K_i},\\
\operatorname{SchoolMin}_i&=\max\left(1, [\frac{n_i^{\mathrm{school}}}{\overline K_i}]-\overline{s}_i^-\right),\\
\operatorname{SchoolMax}_i&=\max\left(\operatorname{SchoolMin}_i, [\frac{n_i^{\mathrm{school}}}{\overline K_i}]+\overline{s}_i^+\right).\\
\end{aligned}
\end{equation}
Centers are deterministically regenerated for the final value of \(\overline K_i\), using a fixed seed. These centers and city-scaled quantities form the initial MILP configuration \(\boldsymbol{\omega}_i^{[0]}\).

\paragraph{Conditional Rescue.} Conditional Rescue is activated only when the initial zoning MILP reports infeasibility or reaches its time limit. It neither retrains the Parameter Generator nor invokes Oracle Search. Instead, it constructs an ordered and bounded sequence of solver-ready repair configurations, 
\begin{equation}
\label{eq:feasibility_aware}
\left(
\boldsymbol{\omega}_i^{[1]},\ldots,\boldsymbol{\omega}_i^{[M_i]}\right)=\operatorname{Rescue}\left(C_i,\overline{\boldsymbol{\theta}}_i,\widehat{\delta}_i^{\mathrm{gen}},\sigma_i\right).
\end{equation}
The number of candidates \(M_i\) depends on the initial solver status and the eligibility conditions satisfied by the city. The complete trigger conditions, parameter modifications, candidate ordering, solver budgets, and stopping rules of the procedure are provided in Algorithm \ref{alg:rescue_procedure}. Only after all applicable candidates have been exhausted without success, we report this city as unresolved.

\begin{algorithm}[h!t]
\caption{Solver-verified Conditional Rescue procedure.}
\label{alg:rescue_procedure}
\noindent\textbf{Input:}
Initial solver-ready configuration \(\boldsymbol{\omega}_i^{[0]}\); city instance \(C_i\); admissible predicted control vector \(\overline{\boldsymbol{\theta}}_i\); predicted plan-generation difficulty \(\widehat{\delta}_i^{\mathrm{gen}}\); failed initial MILP status \(\sigma_i\); fixed policy specification \(\boldsymbol{\rho}\).

\noindent\textbf{Output:}
The first solver-verified successful configuration \(\boldsymbol{\omega}'\) and its zoning plan \(\widehat{\boldsymbol{\pi}}_i^{\prime}\); otherwise, \(\textsc{unresolved}\).
\begin{algorithmic}[1]
\State \textbf{M1: Time-Limit Recovery with Adaptive Zone-Count Increase}
\If{\(\sigma_i=\text{\textsc{time limit}}\)}
    \State \((\sigma_i^{\mathrm{retry}},\widehat{\boldsymbol{\pi}}_i^{[0]}) \gets \Call{MILP}{C_i,\boldsymbol{\omega}_i^{[0]},\boldsymbol{\rho},600~\mathrm{s}}\)

    \State \textbf{if} \(\sigma_i^{\mathrm{retry}}=\text{\textsc{success}}\) \textbf{ then return } \((\boldsymbol{\omega}_i^{[0]},\widehat{\boldsymbol{\pi}}_i^{[0]})\)

    \If{\(\sigma_i^{\mathrm{retry}}=\text{\textsc{time limit}}\)}
        \ForAll{\(\boldsymbol{\omega}'\in\Call{IncreaseZones}{\boldsymbol{\omega}_i^{[0]}}\)}
            \State \((\sigma_i^{\prime},\widehat{\boldsymbol{\pi}}_i^{\prime}) \gets \Call{MILP}{C_i,\boldsymbol{\omega}',\boldsymbol{\rho},600~\mathrm{s}}\)

            \State \textbf{if} \(\sigma_i^{\prime}=\text{\textsc{success}}\) \textbf{ then return } \((\boldsymbol{\omega}',\widehat{\boldsymbol{\pi}}_i^{\prime})\)
        \EndFor
    \EndIf

    \State \Return \(\text{\textsc{unresolved}}\)
\EndIf
\State \textbf{M2: Conditional One-Zone Reconstruction}
\If{
    \(\sigma_i=\text{\textsc{infeasible}}\),
    \(\overline K_i>1\),
    \(n_i^{\mathrm{school}}\geq2\), \textbf{ and }
    \(\bigl(n_i^{\mathrm{school}}\leq3
    \textbf{ or }\Call{SchoolsClustered}{C_i}\bigr)\)
}
    \State \((\boldsymbol{\omega}',\widehat{\boldsymbol{\pi}}_i^{\prime}) \gets \Call{OneZone}{C_i,\boldsymbol{\omega}_i^{[0]}}\)

    \State \textbf{if} \(\Call{Verify}{C_i,\boldsymbol{\omega}',\widehat{\boldsymbol{\pi}}_i^{\prime},\boldsymbol{\rho}}\) \textbf{ then return } \((\boldsymbol{\omega}',\widehat{\boldsymbol{\pi}}_i^{\prime})\) \textbf{ else return } \(\text{\textsc{unresolved}}\)
\EndIf
\State \textbf{M3: Moderate-to-Hard/Extreme Relaxation}
\If{
    \(\sigma_i=\text{\textsc{infeasible}}\),
    \(\widehat{\delta}_i^{\mathrm{gen}}=\mathrm{Moderate}\),
    \(n_i^{\mathrm{school}}\leq10\), \textbf{ and }
    \(n_i^{\mathrm{school}}/\overline K_i\leq3\)
}
    \State \(\boldsymbol{\omega}_i^{H}\gets P_H(\boldsymbol{\omega}_i^{[0]})\)

    \State \((\sigma_i^{H},\widehat{\boldsymbol{\pi}}_i^{H}) \gets \Call{MILP}{C_i,\boldsymbol{\omega}_i^{H},\boldsymbol{\rho},30~\mathrm{s}}\)

    \State \textbf{if} \(\sigma_i^{H}=\text{\textsc{success}}\) \textbf{ then return } \((\boldsymbol{\omega}_i^{H},\widehat{\boldsymbol{\pi}}_i^{H})\)

    \State \textbf{if} \(\sigma_i^{H}\neq\text{\textsc{infeasible}}\) \textbf{ then return } \(\text{\textsc{unresolved}}\)

    \State \(\boldsymbol{\omega}_i^{X}\gets P_X(\boldsymbol{\omega}_i^{[0]})\)

    \State \((\sigma_i^{X},\widehat{\boldsymbol{\pi}}_i^{X}) \gets \Call{MILP}{C_i,\boldsymbol{\omega}_i^{X},\boldsymbol{\rho},30~\mathrm{s}}\)

    \State \textbf{if} \(\sigma_i^{X}=\text{\textsc{success}}\) \textbf{ then return } \((\boldsymbol{\omega}_i^{X},\widehat{\boldsymbol{\pi}}_i^{X})\)

    \State \textbf{if} \(\overline K_i>2\) \textbf{ and } \(n_i^{\mathrm{school}}/\overline K_i<10\) \textbf{ then continue to Method 4; otherwise continue to Method 5}
\EndIf
\State \textbf{M4: Progressive Zone-Count Reduction}
\If{
    \(\sigma_i=\text{\textsc{infeasible}}\),
    \(\overline K_i>2\), \textbf{ and }
    \(n_i^{\mathrm{school}}/\overline K_i<10\)
}
    \ForAll{\(\boldsymbol{\omega}'\in\Call{ReduceZones}{\boldsymbol{\omega}_i^{[0]}}\)}
        \State \((\sigma_i^{\prime},\widehat{\boldsymbol{\pi}}_i^{\prime}) \gets \Call{MILP}{C_i,\boldsymbol{\omega}',\boldsymbol{\rho},30~\mathrm{s}}\)

        \State \textbf{if} \(\sigma_i^{\prime}=\text{\textsc{success}}\) \textbf{ then return } \((\boldsymbol{\omega}',\widehat{\boldsymbol{\pi}}_i^{\prime})\)
    \EndFor

    \State \textbf{continue to Method 5}
\EndIf
\State \textbf{M5: Ordered Final-Grid Relaxation}
\ForAll{\(\boldsymbol{\omega}'\in\Call{FinalGrid}{\boldsymbol{\omega}_i^{[0]}}\)}
    \State \((\sigma_i^{\prime},\widehat{\boldsymbol{\pi}}_i^{\prime}) \gets \Call{MILP}{C_i,\boldsymbol{\omega}',\boldsymbol{\rho},30~\mathrm{s}}\)

    \State \textbf{if} \(\sigma_i^{\prime}=\text{\textsc{success}}\) \textbf{ then return } \((\boldsymbol{\omega}',\widehat{\boldsymbol{\pi}}_i^{\prime})\)
\EndFor

\State \Return \(\text{\textsc{unresolved}}\)
\end{algorithmic}
\end{algorithm}

\clearpage
\endgroup

\end{document}